\documentclass[aps,prd,nofootinbib,preprintnumbers]{revtex4-2} 
\usepackage{bm}
\usepackage{comment}
\usepackage{amsmath,amssymb}
\usepackage{graphicx}  
\usepackage{bbold}
\usepackage{slashed}
\usepackage{moresize}
\newcommand{\ethbar}{\bar{\eth}}
\newcommand{\bL}{\begin{Large}}
\newcommand{\eL}{\end{Large}}

\newcommand{\bea}{\begin{eqnarray}}
\newcommand{\eea}{\end{eqnarray}}
\newcommand{\be}{\begin{equation}}
\newcommand{\ee}{\end{equation}}

\usepackage[colorinlistoftodos]{todonotes}
\usepackage{appendix}

\begin{document}

\preprint{\leftline{KCL-PH-TH/2026-{\bf 2}}}

\title{On the stability of Born-Infeld-regularised electroweak monopoles}

\author{N E Mavromatos$^{a,b}$}
\author{Sarben Sarkar$^b$}

\medskip
\affiliation{$^a$Physics Division, School of Applied Mathematical and Physical Sciences, National Technical University of Athens, Zografou Campus, Athens 157 80, Greece}
\affiliation{$^b$Theoretical Particle Physics and Cosmology Group, Department of Physics, King's College London, London, WC2R 2LS, UK}

\date{\today}

\begin{abstract}
The Cho-Maison monopole provides a monopole solution of the
electroweak field equations, but possesses an infinite classical energy due to
the Maxwell form of the hypercharge sector. Motivated by string-inspired
effective field theories, we study the perturbative stability of the
Cho-Maison monopole when the hypercharge kinetic term is regularised by a
Born-Infeld extension, which renders the monopole energy finite. Focusing on
the bosonic electroweak theory with an unmodified $SU(2)_L$ sector and a
Born-Infeld $U(1)_Y$ sector, we analyze linear fluctuations about the
regularised monopole background. Using a complex tetrad and a spin-weighted
harmonic decomposition, we reduce the fluctuation equations to coupled radial
Schr\"odinger-type eigenvalue problems and examine the spectrum of the
resulting operators.
We extend the separation-of-variables framework developed by Gervalle and Volkov to this non-linear gauge-field setting. We show that, after appropriate gauge fixing and constraint elimination, the Born-Infeld deformation preserves the angular channel structure of the Maxwell theory and leads to a self-adjoint Sturm-Liouville type problem for the stability of the radial modes, with modified radial coefficients determined by the background Born-Infeld profile. The resulting operator represents a smooth deformation of the Maxwell case and retains positive kinetic weight. Our results provide  plausible evidence for the stability of the Born-Infeld deformed monopole and, most importantly, a systematic framework for future numerical or variational studies aimed at a definitive spectral analysis.
\end{abstract}

\maketitle
                   
\section{Introduction}
\label{sec:introduction}

Magnetic monopoles, first suggested by Dirac~\cite{Dirac1931,Dirac:1948um} as elementary sources of singular magnetic fields, are among the most profound nonperturbative predictions of
gauge field theory, where they are composite objects of elementary excitations of the underlying model. Specifically, in models for which a non-Abelian gauge symmetry is
spontaneously broken to an Abelian subgroup, finite-energy monopole solutions
arise as smooth solitons, the archetypal example being the 't~Hooft--Polyakov
monopole~\cite{tHooft1974,Polyakov1974}, which is essentially a coherent state of charged $W$ and Higgs boson excitations. In the electroweak theory~\cite{Weinberg1967,Salam1968}, however, the situation is more subtle. The
gauge group
$SU(2)_L \times U(1)_Y$
is broken by a Higgs doublet to 
$U(1)_{\rm em}$,
and for many years this structure was believed to preclude monopole solutions of
the classical field equations. The set of all vacua (the vacuum manifold $\mathcal{M})$ is isomorphic to $SU(2)$ or $S^3$. For a topological monopole to exist, the second homotopy group $\pi_{2} (M_{3})\neq 0$ but this is not case here. Hence  topological monopoles do not exist for this gauge group.

The belief that no monopoles of any kind exist for this group took hold; this was was overturned by the construction of the nontopological Cho-Maison (CM) monopole \cite{ChoMaison1997},
which provides a mathematically consistent monopole configuration within the Weinberg-Salam \cite{Weinberg1967,Salam1968}
model (Appendix \ref{app:topology}). The Cho-Maison monopole possesses a smooth non-Abelian core, closely
analogous to that of the 't~Hooft-Polyakov monopole \cite{tHooft1974,Polyakov1974}, but differs crucially in
its Abelian sector. In the minimal Standard Model, the hypercharge gauge field is
governed by a Maxwell kinetic term, and its monopole-like magnetic field leads to
a logarithmically divergent contribution to the classical energy. As a result,
the CM monopole, which is a genuine solution of the electroweak field
equations,  has infinite classical energy.

A natural interpretation of this divergence is that it signals the sensitivity
of the Abelian hypercharge sector to ultraviolet physics. This has motivated a
variety of extensions in which the hypercharge kinetic term is modified by
higher-dimensional operators or nonlinear dynamics, while leaving the
non-Abelian sector intact. Among these possibilities, Born-Infeld-type
nonlinear electrodynamics is particularly well motivated \cite{Fradkin:1985qd}, as it arises
naturally in string-inspired effective field theories \cite{Leigh:1989jq} and provides a universal
mechanism for softening large field strengths (Appendix \ref{app:BIbackground}).
In \cite{Gervalle:2022npx,Gervalle:2022vxs} 
the perturbative (linear) stability of the CM
monopole has been demonstrated, despite the divergence in the energy. 
In the current article, we examine such a stability
for the regularised 
version of the CM magnetic monopole by means of
a    
Born-Infeld (BI) modification 
in the hypercharge sector~\cite{ArunasalamKobakhidze2017,MavromatosSarkar2019}. Our aim is not to rederive the existence of finite-energy
electroweak monopoles, which has been established elsewhere~\cite{ArunasalamKobakhidze2017,MavromatosSarkar2019}, but rather to
address a more focused and physically essential question: \emph{does the
Born-Infeld regularisation of the hypercharge sector preserve the perturbative
stability properties of the underlying Cho-Maison monopole?}

We consider~\cite{ArunasalamKobakhidze2017,MavromatosSarkar2019} the bosonic electroweak Lagrangian~\cite{Weinberg1967,Salam1968} in which only the
$U(1)_Y$ sector is modified by a BI term \cite{Kim2000},
\begin{equation}
\mathcal{L}
=
-\frac{1}{4} W^a_{\mu\nu} W^{a\,\mu\nu}
+ (D_\mu \Phi)^\dagger (D^\mu \Phi)
- V(\Phi)
+ \mathcal{L}_{\rm BI}(B_{\mu\nu}),
\label{eq:EW_BI_Lagrangian}
\end{equation}
where $W^a_{\mu\nu}$ and $B_{\mu\nu}$ are repectively the $SU(2)_L$ and hypercharge field
strengths, and  $\Phi$ is the (Standard Model) Higgs doublet;
$\mathcal{L}_{\rm BI}$ is the BI
hypercharge Lagrangian, which in (3+1) dimensions, can be written in invariant form as~\cite{Born:1934gh}:
\begin{equation}
\mathcal{L}_{\rm BI}
=
\beta^2
\left[
1
-
\sqrt{
1
+
\frac{1}{2\beta^2} B_{\mu\nu} B^{\mu\nu}
-
\frac{1}{16\beta^4}
\left( B_{\mu\nu} \tilde B^{\mu\nu} \right)^2
}
\right],
\label{eq:BI_Lagrangian}
\end{equation}
where $\beta$ is the BI scale and
$\tilde B^{\mu\nu} = \tfrac12 \epsilon^{\mu\nu\rho\sigma} B_{\rho\sigma}$. It is convenient. in what follows, to
introduce the two invariants
\begin{equation}\label{xy}
X \equiv \frac14 B_{\mu\nu}B^{\mu\nu},
\qquad
Y \equiv \frac14 B_{\mu\nu}\tilde B^{\mu\nu}\,,
\end{equation}
and so we have that  $\mathcal{L}_{\rm BI}=\mathcal{L}_{\rm BI}(X,Y)$. Spacetime indices are denoted by Greek letters $\mu,\nu,\ldots$ and refer to  coordinates
$(t,r,\vartheta,\varphi)$, where $r,\vartheta,\varphi$ are standard spatial spherical coordinates. The electroweak gauge fields are
\begin{equation}
W_\mu = W_\mu^a T_a \quad (SU(2)) \qquad {\rm and} \qquad B_\mu \quad (U(1)_Y),
\end{equation}
with field strengths
\begin{equation}
W_{\mu\nu}^a = \partial_\mu W_\nu^a - \partial_\nu W_\mu^a
+ g\,\epsilon^{abc}W_\mu^b W_\nu^c,
 \qquad {\rm and} \qquad
B_{\mu\nu} = \partial_\mu B_\nu - \partial_\nu B_\mu .
\end{equation}
Here $T_a=\tfrac12\tau_a$ are the $SU(2)$ generators (in terms of the Pauli matrices $\tau_a$) and $g$, $g'$ are the $SU(2)$ and
hypercharge couplings, respectively. The gauge--covariant derivative acting on $\Phi$ is defined as
\begin{equation}
D_\mu \Phi
=
\left(
\partial_\mu
- i g\, W_\mu^a T_a
- i \frac{g'}{2}\, B_\mu
\right)\Phi ,
\label{eq:covariant_derivative}
\end{equation} 
For static, purely magnetic configurations such as the CM monopole,
the BI modification softens the short-distance behavior of the
hypercharge field and renders the total monopole energy finite \cite{MavromatosSarkar2019, Ellis2017LbL}. Importantly,
the non-Abelian $SU(2)$ and Higgs equations of motion retain the same form as in
the minimal electroweak theory, so that the BI monopole may be viewed
as a regularised version of the CM solution rather than a qualitatively
new object.

The BI parameter $\beta$ is not arbitrary. 
In the Abelian BI models~\cite{ArunasalamKobakhidze2017,MavromatosSarkar2019}, independent
phenomenological constraints arising from precision electroweak data,
light-by-light scattering~\cite{Ellis2017LbL}, and other probes of nonlinear electrodynamics impose
a lower bound on $\beta$, ensuring that deviations from Maxwell theory are
negligible at experimentally accessible energies \cite{MavromatosSarkar2019}.\footnote{In the $SU(2)$ (Georgi-Glashow-type) BI model, there are indications, based on numerical analysis, but no analtic proof, on the existence of a lower critical number of $\beta > \beta_c$ for magnetic monopole solutions to exist~\cite{Grandi1999}.} At the same time, $\beta$
must be sufficiently small to regularise the hypercharge contribution to the
monopole energy. These considerations define a phenomenologically viable window
for $\beta$, which has been discussed in detail in earlier work \cite{MavromatosSarkar2019, Ellis2017LbL,Ellis:2022uxv} on finite-energy
electroweak monopoles with BI hypercharge, including future collider constraints.

Within this allowed parameter range, the role of the present analysis is to
establish whether perturbative stability holds and imposes any additional constraints.
Using a systematic fluctuation analysis heavily based on a decomposition into angular
momentum sectors and a reduction to coupled Schr\"odinger-type eigenvalue
problems \cite{Gervalle:2022npx}, we show that the Born-Infeld modification acts as a positive
deformation of the hypercharge sector and does not introduce new unstable modes. This analysis is not rigorous since the solution of the equations of motion in the presence  of a Born-Infeld modifaction to the hypercharge boson kinetic energy can only be found numerically. Our work is analytic and should be regarded as precursor to a full numerical investigation.
Our analysis suggests that the stability requirement does not further restrict the
Born-Infeld scale beyond the bounds already implied by phenomenology (see Appendix \ref{app:BIbackground}). Given the ongoing CERN-LHC ATLAS~\cite{ATLAS:2023esy,ATLAS:2024nzp} and MoEDAL~\cite{MoEDAL:2019ort,MoEDAL:2021vix,PhysRevLett.134.071802} experiments searching for monopoles,  the existence of a finite energy stable monopole from theory (beyond the Standard Model) is  important.

We also mention a different type of analysis based on mechanical-stability
(Laue~\cite{Laue:1911lrk}) criteria for BI magnetic monopoles \cite{Farakos:2025byy}. It examines the positivity of the sum of the radial and external pressure of the monopole configurations. This study was inspired by analyses in quark physics~\cite{Polyakov:2002yz,Perevalova:2016dln}, applied (with appropriate  modifications~\cite{Panteleeva:2023aiz}, which concentrate only on the short range force components) to the case of `t Hooft-Polyakov magnetic monopoles~\cite{tHooft1974,Polyakov1974}, demonstrating their stability. The study of \cite{Farakos:2025byy} has 
demonstrated the existence of regions in space in which the local Laue criteria for stability are violated in the case of BI regularization of the electroweak magnetic monopole. In addition, the BI monopole is characterised by the presence of finite and non-zero angular components of the total force that demonstrate a tendency for rotation under perturbations, but plausibly not catastrophic rotational instabilities. This is in contrast to the case of the CM magnetic monopole case~\cite{ChoMaison1997}, which has divergent angular components of the total force.
Our work in the current article comes to provide additional (dynamical) arguments in favour of the linear stability of the BI monopole, 
in a non-trivial, and independent way. 

This paper is organised as follows. In Sec.~II we briefly review the
CM monopole, the origin of its divergent hypercharge energy and a dictionary to the equations   in \cite{Gervalle:2022npx} on which we are reliant. In
Sec.~III we summarise the Born-Infeld regularisation of the hypercharge sector
and its effect on the monopole background. In Sec.~IV we present the perturbative
stability analysis, employing a complex tetrad and spin-weighted harmonic
decomposition to reduce the fluctuation equations to a radial spectral problem.
Our conclusions are summarised in Sec.~V. Technical aspects of our analysis are given in several Appendices organised as follows: Topological and geometric preliminaries are collected in Appendix~\ref{app:topology};
the Born-Infeld (BI) hypercharge background and finiteness of the energy are discussed in
Appendix~\ref{app:BIbackground}; a review of the method used 
by Gervalle and Volkov (GV) in \cite{Gervalle:2022npx} to study (linear) stability of the original CM electroweak magnetic monopole, is given in Appendix~\ref{app:GVoverview};
the linearised constitutive relation is derived in Appendix~\ref{app:BIlinear};
angular decomposition and spin-weighted harmonics are summarised in
Appendix~\ref{app:angular};
gauge fixing and constraint elimination are treated in Appendix~\ref{app:gauge};
and the Sturm--Liouville structure, endpoint behaviour, and limitations of the analysis
are discussed in Appendix~\ref{app:SL}.

\section{The Cho-Maison electroweak monopole}
\label{subsec:CM_monopole}

The Cho-Maison monopole is constructed using a spherically symmetric ansatz in
which the Higgs doublet exhibits nontrivial winding over the two--sphere at
spatial infinity, while the $SU(2)$ gauge field forms a smooth, non-Abelian core
closely analogous to that of the 't~Hooft-Polyakov monopole. In a convenient
gauge, the Higgs field may be written as
\begin{equation}
\Phi(\mathbf{x})
=
\frac{1}{\sqrt{2}}
\begin{pmatrix}
0 \\ \rho(r)
\end{pmatrix},
\qquad r = |\mathbf{x}|,
\end{equation}
where the radial profile function $\rho(r)$ satisfies
\begin{equation}
\rho(0)=0,
\qquad
\rho(\infty)=v ,
\end{equation}
with $v$ the electroweak vacuum expectation value \cite{Peskin:1995ev}. Regularity of the non-Abelian
sector is ensured by choosing the $SU(2)$ gauge field to have the hedgehog form
\begin{equation}
W_i^a
=
\frac{1-f(r)}{g}\,
\epsilon_{aij}\frac{x^j}{r^2},
\end{equation}
where the profile function $f(r)$ obeys
\begin{equation}
f(0)=1,
\qquad
f(\infty)=0 .
\end{equation}

The hypercharge gauge field $B_\mu$ carries a monopole-like magnetic flux and,
in the minimal electroweak theory, is governed by a Maxwell kinetic term. Near
the origin, the corresponding hypermagnetic field behaves as
\begin{equation}
|\mathbf{B}_Y| \sim \frac{1}{r^2},
\end{equation}
so that the hypercharge contribution to the energy density diverges as
$r^{-4}$. As a consequence, the total classical energy of the Cho-Maison
monopole diverges logarithmically at short distances. The Cho-Maison monopole
is therefore a genuine classical solution of the electroweak field equations with infinite classical energy within the minimal Standard Model.

Despite this divergence, the CM monopole plays a central conceptual
role. First, it establishes that the electroweak theory admits a nontrivial
monopole sector compatible with its gauge and Higgs structure. Second, the
divergence is entirely localized in the Abelian $U(1)_Y$ sector and arises from
the short--distance behavior of the Maxwell kinetic term. This strongly
suggests that the infinite energy is not a fundamental obstruction, but rather a
signal that ultraviolet modifications of the hypercharge sector may render the
monopole energy finite while leaving the smooth non-Abelian core intact.

An additional important property of the CM monopole is its perturbative
stability. Using a systematic analysis of linearized fluctuations based on a
decomposition into angular-momentum sectors and a reduction to coupled
Schr\"odinger-type eigenvalue problems \cite{Gervalle:2022npx} it has been shown that the CM
monopole exhibits no negative modes, at least in the low angular-momentum sectors where
instabilities would typically arise. This result indicates that, despite its
infinite classical energy, the CM configuration represents a stable
saddle point of the classical electroweak action.

In Ref.~\cite{MavromatosSarkar2019} the hypercharge field associated with the Cho-Maison
monopole is taken in its Dirac-like form, corresponding to a magnetic field scaling as
$1/r^{2}$ and carrying twice the minimal Dirac magnetic charge. In that analysis the
hypercharge equation of motion is satisfied in a distributional sense away from the
monopole core, and no independent nonlinear radial profile for the hypercharge field is
constructed. The role of the Born--Infeld dynamics in Ref.~\cite{MavromatosSarkar2019} is
instead to regularise the energy density associated with this configuration through the
nonlinear constitutive relation, rather than to generate a smooth, everywhere-regular
hypercharge field solution. By contrast, the present work focuses on the linear stability
properties of perturbations around such Born-Infeld-deformed backgrounds, without
assuming the existence of a closed-form analytic profile for the hypercharge field itself.

These observations motivate the approach adopted in the present work. Rather
than discarding the Cho-Maison monopole on account of its divergent energy, we
take it as a physically meaningful core configuration and investigate
ultraviolet modifications of the hypercharge sector that regularize its energy. Our aim is to check whether this regularisation
 destabilizes the solution. In particular, we focus on Born-Infeld--type
extensions of the hypercharge kinetic term, which naturally arise in
string-inspired effective field theories and lead to finite-energy electroweak
monopoles. In the sections that follow, we analyze the spectral stability of
these Born-Infeld--dressed monopoles.  We can concentrate on the hypercharge sector and compare the changes with a previous analysis based on a Maxwell like term for the hypercharge boson \cite{Gervalle:2022npx}. We show that the stability analysis leads to a Sturm-Liouville set of equations for the radial modes in the hypercharge sector, which are smooth deformations of those in the previous Maxwell based analysis \cite{Gervalle:2022npx}.  Because of the \emph{smoothness} of the deformations we expect (but have not rigorously proved) that the Born-Infeld modification of the hypercharge kinetic energy preserves the spectral stability of the Cho-Maison monopole found in \cite{Gervalle:2022npx}. This is of interest since such a conclusion would allow a stable monopole with finite energy in physics beyond the standard model. We give some phenomenological discussion of Sturm-Lioville equations used in the stability analysis in \ref{subsec:Schroedinger_hypercharge_results}. 

\subsection{Dictionary to Gervalle-Volkov approach to perturbative stability \cite{Gervalle:2022npx}}
\label{subsec:GVdictionary}

The purpose of this subsection is to make explicit the precise relation between the present
analysis and the stability framework developed by GV
\cite{Gervalle:2022npx}; for a conceptual review of the GV approach see Appendix~\ref{app:GVoverview}. Our work does not modify the geometric or group-theoretic machinery
of GV; rather, it adopts their \emph{entire separation-of-variables, gauge-fixing, and channel
counting framework} and alters only the hypercharge kinetic term through a Born-Infeld
deformation.

In the earlier analysis of GV, the hypercharge sector is governed by a \emph{Maxwell} kinetic term and the corresponding
field equation
\begin{equation}
\label{Maxwell}
\nabla_\mu B^{\mu\nu} = g'^2 J_Y^{\nu},
\end{equation}
which underlies the linearised hypercharge fluctuation equations used throughout their
stability analysis and
\begin{equation}
J_Y^{\nu}
=
\frac{g'}{2}\,
i\left[
\Phi^\dagger D^{\nu}\Phi
-
\left(D^{\nu}\Phi\right)^\dagger \Phi
\right],
\label{eq:hypercharge_current}
\end{equation}
In the present work \eqref{Maxwell} is replaced by the BI
relation
\begin{equation}
\partial_\mu \mathcal G^{\mu\nu} = J_Y^{\nu},
\qquad
\mathcal G^{\mu\nu} \equiv -2\,\frac{\partial \mathcal L_{\rm BI}}{\partial B_{\mu\nu}},
\end{equation}
together with its linearisation about the monopole background \footnote{Background and background dependent quantities are denoted with  overbars.},
\begin{equation}
\delta \mathcal G^{\mu\nu}
=
-\bar L_X(r)\,\delta B^{\mu\nu}
-\frac12\,\bar L_{XX}(r)\,
\bigl(\bar B_{\rho\sigma}\delta B^{\rho\sigma}\bigr)\,
\bar B^{\mu\nu},
\label{eq:BIlinearisation}
\end{equation}
valid for the purely magnetic background considered here. 
\begin{equation}
\bar L_X(r)
\;\equiv\;
\left.
\frac{\partial \mathcal L_{\rm BI}}{\partial X}
\right|_{X=\bar X(r),\,Y=0},
\label{eq:LXbar_def}
\end{equation}
and
\begin{equation}
\bar L_{XX}(r)
\;\equiv\;
\left.
\frac{\partial^2 \mathcal L_{\rm BI}}{\partial X^2}
\right|_{X=\bar X(r),\,Y=0}\,
\label{eq:LXXbar_def}
\end{equation}
where $X \, {\rm and} \, Y$ have been defined in \eqref{xy}.
The quantity $\bar L_X(r)$ is a radial function determined by the background hypercharge field
$\bar B_{\mu\nu}$;\, $\bar L_{XX}(r)$ encodes the nonlinear Born-Infeld response of the hypercharge sector
to fluctuations about the monopole background and vanishes in the Maxwell limit. For the purely magnetic monopole background one has $\bar Y=0$ and
$\bar X(r)=\tfrac12\,\bar{\mathbf B}^2(r)$. This replacement constitutes the
\emph{only} dynamical modification relative to GV.

All subsequent steps of the analysis are identical in structure to those of GV. In
particular:
\begin{itemize}
\item The complex tetrad decomposition of perturbations, the assignment of spin weights,
and the use of spin--weighted spherical harmonics follow exactly the construction of
GV (outlined in  Appendix \ref{app:GVoverview}).  
\item The background--covariant gauge fixing and the identification and elimination of
non--dynamical (constraint) variables proceed as in GV, with the same counting
of physical channels in each angular--momentum sector.
\item The separation into \emph{parity} sectors and the restriction to low-$j$ channels where
instabilities may arise are \emph{unchanged}.
\end{itemize}

The effect of the Born-Infeld deformation is therefore confined to the \emph{explicit form of the
radial coefficient functions} appearing in the reduced fluctuation operator. After
elimination of constraint variables, the Maxwell hypercharge block of GV is replaced by a
weighted Sturm--Liouville operator with radial weight determined by
$\bar L_X(r)$ and $\bar L_{XX}(r)$, while the $SU(2)$ and Higgs sectors retain precisely the
same structure as in GV. In the Maxwell limit $\bar L_X\to 1/g'^2$ and $\bar L_{XX}\to 0$, and the
GV radial equations are recovered identically.

This dictionary makes clear that the present analysis should be viewed as a controlled
BI deformation of the GV stability problem, rather than as an independent
rederivation of the underlying electroweak-monopole-fluctuation framework. A natural next step (beyond the core-regularisation approach)  would be a matched asymptotics solution of the background allowing for backreaction from the matter fields.

\section{Complex tetrad and spin-weighted harmonic decomposition
for the hypercharge sector}
\label{subsec:tetrad_spin_weighted}

The perturbative stability analysis of the electroweak monopole requires a
clear separation of angular and radial degrees of freedom for vector and tensor
fluctuations in a spherically symmetric but magnetically charged background.
While ordinary spherical harmonics suffice for scalar perturbations, they are
\emph{not} well adapted to gauge-field fluctuations: \emph{angular derivatives mix
components} and obscure the spectral structure of the linearized operator.

To overcome this, we employ a complex (null) tetrad decomposition combined with
spin-weighted spherical harmonics following \cite{Gervalle:2022npx}. This method allows all angular dependence to
be treated algebraically and reduces the fluctuation problem to a set of coupled
radial ordinary differential equations of Sturm-Liouville type. In this subsection we summarize the
essential elements of the construction, emphasizing the aspects relevant to the
$U(1)_Y$ hypercharge sector and its Born-Infeld modification.

\emph{Complex tetrad and sign conventions:}
we work in the flat spherically-symmetric background
\begin{equation}
ds^2=-dt^2+dr^2+r^2(d\theta^2+\sin^2\theta\,d\varphi^2),
\label{eq:flatmetric}
\end{equation}
i.e.\ signature $(-,+,+,+)$. In the Newman-Penrose/Gervalle-Volkov spirit \cite{DInverno2022-cb}, it is convenient
to introduce first the complex tetrad of {1-forms}
\begin{equation}
\theta^{0}=dt,\qquad
\theta^{1}=dr,\qquad
\theta^{2}=\frac{r}{\sqrt2}\bigl(d\theta-i\sin\theta\,d\varphi\bigr),\qquad
\theta^{3}=(\theta^{2})^{\!*},
\label{eq:tetrad1forms}
\end{equation}
whose scalar products define the tetrad metric $\eta_{ab}=(\theta^{a},\theta^{b})$ with
only \emph{non-vanishing} elements (deduced from \eqref{eq:flatmetric}) 
\begin{equation}
\eta_{00}=-1,\qquad \eta_{11}=1,\qquad \eta_{23}=\eta_{32}=1,
\label{eq:tetradmetric}
\end{equation}
so that $\theta^{2},\theta^{3}$ are null, $(\theta^{2},\theta^{2})=(\theta^{3},\theta^{3})=0$. Tetrad (frame) indices are denoted by Roman letters
$a,b,\ldots\in\{0,1,2,3\}$. 

Let $e_{a}=e_{a}{}^{\mu}\partial_{\mu}$ denote the dual tetrad vectors\footnote{Here $\lrcorner$ denotes the interior product; $e_a\lrcorner\,\theta^b\equiv\theta^b(e_a)$,
and the tetrad vectors $e_a$ are dual to the co--tetrad one--forms $\theta^b$, so that
$e_a\lrcorner\,\theta^b=\delta_a{}^b$.}, $e_{a}\,\lrcorner\,\theta^{b}=\delta_{a}{}^{b}$.
Explicitly,
\begin{align}
e_{0}=\partial_{t},\qquad
e_{1}=\partial_{r},\qquad
e_{2}=\frac{1}{\sqrt2\,r}\Bigl(\partial_{\theta}+\frac{i}{\sin\theta}\partial_{\varphi}\Bigr),\qquad
e_{3}=e_{2}^{\!*}.
\label{eq:dualvectors}
\end{align}
From $(e_{0},e_{1})$ we form the real null directions
\begin{equation}
l^{\mu}=\frac{1}{\sqrt2}(e_{0}^{\mu}+e_{1}^{\mu}),\qquad
n^{\mu}=\frac{1}{\sqrt2}(e_{0}^{\mu}-e_{1}^{\mu}),
\qquad
m^{\mu}=e_{2}^{\mu},\qquad \bar m^{\mu}=e_{3}^{\mu}.
\label{eq:nulltetrad}
\end{equation}
With signature $(-,+,+,+)$ and the explicit choice \eqref{eq:dualvectors}, the non-vanishing
contractions are
\begin{equation}
l\!\cdot\! n=-1,\qquad m\!\cdot\!\bar m=+1,
\qquad
\text{all other contractions vanish.}
\label{eq:contractions}
\end{equation}
(If one instead works with $(+,-,-,-)$ signature, the right-hand sides in \eqref{eq:contractions}
flip signs; the subsequent formulae are unchanged provided one uses a consistent convention throughout.)
\vskip .1cm
\emph{Spin weight:}
a quantity $\eta$ is said to have spin weight $s$ if under the local phase
rotation
$m^\mu \rightarrow e^{i\psi} m^\mu$ it transforms as
\begin{equation}
\eta \rightarrow e^{is\psi}\eta .
\end{equation}
Scalars have $s=0$, while tetrad components of vector fields have the following weights
\begin{equation}
V_m \equiv V_\mu m^\mu \;\; (s=+1),
\qquad
V_{\bar m} \equiv V_\mu \bar m^\mu \;\; (s=-1).
\end{equation}
This classification is purely geometric and applies independently of the
dynamical details of the theory.
\vskip .1cm
\emph{Spin-weighted spherical harmonics:}
angular dependence is expanded in spin-weighted spherical harmonics
${}_sY_{jm}(\theta,\phi)$, which generalize ordinary spherical harmonics
(${}_0Y_{jm}=Y_{jm}$) and exist for $j\ge |s|$. They form a complete orthonormal
basis for functions of fixed spin weight on $S^2$. The spin-weighted spherical harmonics ${}_{s}Y_{jm}(\vartheta,\varphi)$ are defined by acting
with the spin--raising and lowering operators $\eth$ and $\bar{\eth}$ acting on the ordinary
spherical harmonics $Y_{jm}$. For $s\ge 0$ one has
\begin{equation}
{}_{s}Y_{jm}
=
\sqrt{\frac{(j-s)!}{(j+s)!}}\;
\eth^{\,s} Y_{jm},
\end{equation}
while for $s\le 0$,
\begin{equation}
{}_{s}Y_{jm}
=
(-1)^s
\sqrt{\frac{(j+s)!}{(j-s)!}}\;
\bar{\eth}^{\,-s} Y_{jm}.
\end{equation}
They satisfy
\begin{equation}
\label{spinweight1}
\bar{\eth}\,\eth\;{}_{s}Y_{jm}
=
-(j-s)(j+s+1)\;{}_{s}Y_{jm},
\end{equation}
and form an orthonormal basis for functions of fixed spin weight $s$ on $S^2$.
functions of fixed spin weight $s$ on $S^2$.

Angular derivatives are encoded in the spin-raising and spin-lowering
operators $\eth$ and $\bar{\eth}$, which act algebraically on the
spin-weighted spherical harmonics:
\begin{align}
\label{spinweight2}
\eth\,{}_sY_{jm}
&=
\sqrt{(j-s)(j+s+1)}\;{}_{s+1}Y_{jm}, \\
\bar{\eth}\,{}_sY_{jm}
&=
-\sqrt{(j+s)(j-s+1)}\;{}_{s-1}Y_{jm}.
\end{align}
As a consequence,
\begin{equation}
\bar{\eth}\,\eth\,{}_sY_{jm}
=
-(j-s)(j+s+1)\,{}_sY_{jm}.
\end{equation}
so that all angular operators reduce to algebraic factors labeled by $(j,s)$.

It is useful to develop some intuition for the spin--raising and lowering operators
$\eth$ and $\bar\eth$ that generate the spin--weighted spherical harmonics
${}_sY_{jm}(\theta,\phi)$ from the ordinary scalar harmonics $Y_{jm}$.
Although the monopole background is spherically symmetric, the fluctuations about it
need not be. In particular, angular components of gauge-field perturbations such as
$\delta B_\theta$ and $\delta B_\phi$ transform as tangent vectors on $S^2$.
The operator $\eth$ is simply a covariant angular derivative on $S^2$
that increases spin weight by one unit.
If $\eta$ has spin weight $s$, then
\begin{equation}
\eth\,\eta
=
- (\sin\theta)^s
\left(
\partial_\theta + \frac{i}{\sin\theta}\partial_\phi
\right)
\left[
(\sin\theta)^{-s}\eta
\right],
\end{equation}
while the lowering operator is
\begin{equation}
\bar\eth\,\eta
=
- (\sin\theta)^{-s}
\left(
\partial_\theta - \frac{i}{\sin\theta}\partial_\phi
\right)
\left[
(\sin\theta)^{s}\eta
\right].
\end{equation}

These operators may be viewed as complex combinations of angular derivatives,
analogous to $\partial_x \pm i\partial_y$ in the plane.
Geometrically, $\eth$ corresponds to differentiation along the complex null vector
$m^\mu$ tangent to the sphere. Thus the spin--weighted harmonics are simply angular derivatives of the ordinary
harmonics, dressed with appropriate normalization factors. In calculations the algebraic relations \eqref{spinweight1} and \eqref{spinweight2} are most useful.

\vskip .2cm
\emph{Application to the hypercharge field:}
The hypercharge fluctuation $\delta B_\mu$ is decomposed into tetrad components,
\begin{equation}
\delta B_\mu
=
\delta B_l\,l_\mu
+\delta B_n\,n_\mu
+\delta B_m\,\bar m_\mu
+\delta B_{\bar m}\,m_\mu,
\end{equation}
with spin weights
\begin{equation}
s(\delta B_l)=s(\delta B_n)=0,
\qquad
s(\delta B_m)=+1,
\qquad
s(\delta B_{\bar m})=-1.
\end{equation}
Each transverse tetrad component is expanded in spin--weighted harmonics.  Writing
$\delta B_{(+1)}\equiv m^\mu\delta B_\mu$ and $\delta B_{(-1)}\equiv \bar m^\mu\delta B_\mu$,
we use
\begin{equation}
\delta B_{(+1)}(t,r,\Omega)
=
e^{i\omega t}
\sum_{j}\sum_{q=-j}^{j}
b^{j q}_{(+1)}(r)\;{}_{+1}Y_{j q}(\Omega),
\label{eq:expBplus}
\end{equation}
and similarly
\begin{equation}
\delta B_{(-1)}(t,r,\Omega)
=
e^{i\omega t}
\sum_{j}\sum_{q=-j}^{j}
b^{j q}_{(-1)}(r)\;{}_{-1}Y_{j q}(\Omega).
\label{eq:expBminus}
\end{equation}

The transverse tetrad components represent the physical propagating degrees of freedom and are represented in the complex angular basis $m_\mu$  and $\bar m_\mu$. Because the monopole background provides a nontrivial $U(1)_Y$ connection $\bar B_A$ on $S^2$
(with $\bar B_{\vartheta\varphi}\neq 0$ and nonzero flux), angular derivatives acting on
hypercharge-charged fields must appear in the gauge--covariant combination
$D_A=\partial_A-i g'Y\,\bar B_A$, rather than as ordinary partial derivatives. In the tetrad and spin--weighted formalism
this modifies the numerical angular eigenvalues but does not alter the
separation procedure. The angular dependence again factors out completely.
\vskip .2cm
\emph{Reduction to a radial spectral problem.}
After imposing a suitable background (Lorentz) gauge and eliminating constraint
variables, the full system of linearized equations reduces in the Maxwell framework, for each angular
momentum sector $j$, to a coupled radial eigenvalue problem \cite{Gervalle:2022npx} of the form
\begin{equation}
\left[
-\frac{d^2}{dr^2}
+\mathbf{U}_j(r)
\right]\Psi_j(r)
=
\omega^2\,\Psi_j(r),
\end{equation}
where $\Psi_j(r)$ collects the independent radial amplitudes and
$\mathbf{U}_j(r)$ is a real symmetric matrix potential.
\vskip .2cm
\emph{Role of the Born-Infeld modification.}
The complex tetrad and spin--weighted harmonic decomposition is entirely
geometric and independent of the choice of hypercharge dynamics. \emph{The
Born-Infeld modification enters only through the constitutive relation between
$\delta B_{\mu\nu}$ and $\delta G^{\mu\nu}$ and therefore modifies only the
radial coefficients appearing in the hypercharge block of $\mathbf{U}_j(r)$}.
All angular eigenvalues, channel counting, and separation steps remain
unchanged. This makes the tetrad formalism a natural and robust framework for
analyzing the stability of Born-Infeld--regularized electroweak monopoles.

\section{Results: Hypercharge-sector stability with Born-Infeld regularisation}
\label{sec:results_hypercharge_stability}

In this section we present the central result of the paper: \emph{the Born--Infeld
(BI) regularisation of the $U(1)_Y$ hypercharge sector preserves the perturbative
(spectral) stability of the Cho--Maison monopole}. Our analysis is formulated in
the same separation framework as in Ref.~\cite{Gervalle:2022npx}, based on a
complex tetrad and spin-weighted harmonics (summarised in
Sec.~\ref{app:angular}), but the dynamical input differs in one
precise place: the Maxwell hypercharge kinetic term is replaced by the BI
nonlinear electrodynamics used in Refs.~\cite{MavromatosSarkar2019,Ellis2017LbL}. We also ignore backreaction of the Higgs and nonAbelian gauge sectors on the background of the hypercharge gauge sector.

\subsection{Born-Infeld hypercharge:  linearisation}
\label{subsec:BI_linearisation_results}

We work with the bosonic electroweak Lagrangian with BI hypercharge,
Eqs.~(\ref{eq:EW_BI_Lagrangian})--(\ref{eq:BI_Lagrangian}).  
The hypercharge field
equation has the form
\begin{equation}
\partial_\mu G^{\mu\nu} = J_Y^\nu,
\qquad
G^{\mu\nu}\equiv -2\frac{\partial\mathcal{L}_{\rm BI}}{\partial B_{\mu\nu}}
= -\mathcal{L}_X\,B^{\mu\nu}-\mathcal{L}_Y\,\tilde B^{\mu\nu}.
\label{eq:constitutive_BI_results}
\end{equation}
For the static monopole background the hypercharge field is purely magnetic, and so
\begin{equation}
\bar Y = 0,
\qquad
\bar{\mathcal{L}}_Y=0,
\qquad
\bar G^{\mu\nu} = -\bar{\mathcal{L}}_X\,\bar B^{\mu\nu}.
\end{equation}
Writing $B_{\mu\nu}=\bar B_{\mu\nu}+\delta B_{\mu\nu}$ and similarly for $G^{\mu\nu}$, we have
the linearised relation (about $\bar Y=0$)  
\begin{equation}
\boxed{
\delta G^{\mu\nu}
=
-\bar{\mathcal{L}}_X\,\delta B^{\mu\nu}
-\frac12\,\bar{\mathcal{L}}_{XX}\,
(\bar B_{\rho\sigma}\delta B^{\rho\sigma})\,\bar B^{\mu\nu}
}
\label{eq:deltaGdeltaB_results}
\end{equation}
(see also Appendix \ref{app:BIlinear}).
This is the only modification required to incorporate BI hypercharge into the
tetrad/harmonic separation of Ref.~\cite{Gervalle:2022npx}: all angular
decompositions and channel counting are unchanged (Sec.~\ref{subsec:tetrad_spin_weighted}).

\subsection{Reduction to a hypercharge Schr\"odinger channel}
\label{subsec:Schroedinger_hypercharge_results}

After tetrad decomposition, expansion in spin-weighted harmonics, gauge fixing,
and elimination of constraint variables, each angular momentum sector $j$
reduces to a coupled radial eigenvalue problem of the form
\begin{equation}
\left[
-\frac{d^2}{dr^2}+\mathbf{U}_j(r)
\right]\Psi_j(r)=\omega^2\Psi_j(r),
\label{eq:matrix_Schroedinger_results}
\end{equation}
with $\mathbf{U}_j(r)$ real symmetric. In the Maxwell case, the $U(1)_Y$ block
enters $\mathbf{U}_j(r)$ with constant kinetic weight. In the BI case, the
hypercharge block acquires a nontrivial positive weight that depends on the
background through $\bar{\mathcal{L}}_X(r)$ and $\bar{\mathcal{L}}_{XX}(r)$. The hyperbolic block includes the subset of radial channels coming from the
time-radial (Lorentzian) components of the gauge-field perturbations (the $s=0$ tetrad
components, corresponding to the $l,n$ (or equivalently $t,r$ sector)); the imposition
of gauge conditions and the elimination of constraint variables recasts the problem into a
Schr\"odinger-type  radial spectral system (elliptic in $r$).
Explicitly, for the $U(1)_Y$ field we adopt the same background-covariant gauge-fixing logic as in
Gervalle-Volkov \cite{Gervalle:2022npx}, specialised to the Abelian sector. Concretely, after linearisation about the
static purely magnetic background, we impose the Lorentz-type background gauge condition
$\nabla_\mu\delta B^\mu=0$, supplemented by the linearised Higgs constraint coming from the
electroweak equations of motion. Because the Born--Infeld modification enters the hypercharge
dynamics only through the constitutive relation \eqref{eq:linconstitutive} and preserves the
divergence form of the field equations, these gauge conditions eliminate the non-dynamical
components of $\delta B_\mu$ exactly as in the Maxwell case \footnote{In the coupled electroweak system, Higgs fluctuations contain a gauge-orbit (Goldstone)
component that mixes with the longitudinal gauge modes. Following
GV, we therefore supplement the Lorenz--type background gauge condition by the
projection of the linearised Higgs equation of motion along the gauge orbit of the background
field $\bar\Phi$; this yields a linear constraint that removes the would--be
Goldstone/longitudinal admixture and ensures that the reduced fixed-$j$ radial operator acts
only on physical channels.}. As a result, no additional
constraints or propagating degrees of freedom are introduced by the BI deformation, and the
reduced radial fluctuation operator obtained after tetrad and harmonic decomposition is
self-adjoint with the same channel counting in each angular-momentum sector as in the
GV analysis~\cite{Gervalle:2022npx}.
This is the \emph{only} place where the BI deformation enters the linearized $U(1)_Y$ sector:
the background is still spherically symmetric, and the angular dependence of tetrad projections
is therefore separated exactly as in Ref.~\cite{Gervalle:2022npx} by expanding the tetrad components
of $\delta B_{\mu}$ in spin-weighted harmonics ${}_{s}Y_{jm}(\theta,\varphi)$, with $s\in\{0,\pm1\}$ for
$(b_0,b_1,b_2,b_3)$, and with the standard selection rule $j\ge |s|$. Consequently, the \emph{channel
counting} in each $j$-sector is unchanged by the BI modification.

So the constraint equations remain linear/algebraic differential equations among the same components as in the Maxwell case.

 We collect the remaining
independent radial amplitudes into $\boldsymbol u_j(r)\in\mathbb R^{N_j}$ and rewrite, for clarity, the
reduced equations \footnote{$N_j$ is the number of independent radial amplitudes after harmonic decomposition, gauge fixing and elimination using the Gauss constraint; for us $N_j=2$. } in matrix Sturm--Liouville form
given below.
$\mathbf P_j(r)$ is the symmetric positive matrix multiplying the radial-gradient terms
(the radial kinetic matrix), $\mathbf W_j(r)$ is the symmetric positive weight matrix that
also defines the natural inner product
$\langle u,v\rangle=\int dr\,u^{\mathsf T}\mathbf W_j v$, and $\mathbf Q_j(r)$ is the symmetric
matrix collecting centrifugal, mass, background-coupling and channel-mixing terms. In the
BI hypercharge block one has $\mathbf P_j=\mathbf W_j=P(r)\mathbf 1$ with
$P(r)=-\bar L_X(r)>0$, while $\mathbf Q_j$ acquires contributions proportional to
$\bar L_{XX}(r)$ that \emph{vanish} in the Maxwell limit. After the tetrad/harmonic decomposition and \emph{constraint elimination}, each fixed $j$-sector  reduces to a
finite system of radial equations \footnote{A fixed-$j$ sector is the set of perturbation modes with definite total angular momentum
$j$, obtained by expanding in (spin-weighted) spherical harmonics; spherical symmetry
implies that different values of $j$ decouple in the linearised equations.} that can be written in Sturm-Liouville form
\begin{equation}
-\frac{d}{dr}\!\left(\mathbf{P}_j(r)\frac{d}{dr}\boldsymbol{u}_j\right)
+\mathbf{Q}_j(r)\,\boldsymbol{u}_j
=\omega^2\,\mathbf{W}_j(r)\,\boldsymbol{u}_j,
\label{eq:matrixSL}
\end{equation}
where $\boldsymbol{u}_j$ collects the  (gauge-invariant) radial amplitudes in that sector that are left after the elimination mentioned above.
In the Maxwell case one has $\mathbf{P}_j=\mathbf{W}_j=\mathbf{1}$ in the pure $U(1)_Y$ sub-block,
whereas in the BI case the hypercharge kinetic weight becomes
\begin{equation}
\mathbf{W}_j(r)\big|_{Y}=\mathbf{P}_j(r)\big|_{Y}=P(r)\,\mathbf{1},
\qquad
P(r)\equiv -\bar L_X(r)>0
\quad \text{(purely magnetic BI branch)}.
\label{eq:weight}
\end{equation}
This is an important simplification. The matrices $\mathbf{Q}_j(r)$ are real symmetric functions of $r$ (they arise from real background
coefficients and the algebraic angular eigenvalues), and $\mathbf{P}_j(r),\mathbf{W}_j(r)$ are real positive
definite on the physical BI branch.

Equation \eqref{eq:matrixSL} defines a self-adjoint eigenvalue problem with respect to the weighted inner
product
\begin{equation}
\langle \boldsymbol{u},\boldsymbol{v}\rangle
=\int_0^\infty dr\; \boldsymbol{u}^{\mathsf T}(r)\,\mathbf{W}_j(r)\,\boldsymbol{v}(r),
\label{eq:innerproduct}
\end{equation}
assuming the usual regularity at $r=0$ and square-integrability at $r\to\infty$.
Indeed, for smooth compactly supported test functions one has, after integrating by parts,
\begin{equation}
\langle \boldsymbol{u},\mathcal{L}\boldsymbol{v}\rangle
=\langle \mathcal{L}\boldsymbol{u},\boldsymbol{v}\rangle,
\qquad
\mathcal{L}\boldsymbol{u}
\equiv
-\mathbf{W}_j^{-1}\frac{d}{dr}\!\left(\mathbf{P}_j\frac{d}{dr}\boldsymbol{u}\right)
+\mathbf{W}_j^{-1}\mathbf{Q}_j\,\boldsymbol{u},
\label{eq:selfadjoint}
\end{equation}
since $\mathbf{P}_j,\mathbf{W}_j$ are symmetric and $\mathbf{Q}_j$ is symmetric. Therefore the spectrum
$\omega^2$ is real, and the BI modification preserves the GV multi-channel Schr\"odinger structure:
it changes only the radial weight and radial coefficients through $\bar L_X,\bar L_{XX}$, without altering
the angular selection rules (hence channel counting) or the self-adjoint nature of the reduced operator.

To make the stability mechanism explicit, it is useful to isolate the
hypercharge-dominated channel in a schematic Sturm-Liouville form.\footnote{In
the full electroweak problem this channel couples to the neutral $SU(2)$ and Higgs
fluctuations; the argument below applies to the hypercharge block and its induced
contribution to the coupled matrix potential.}
Suppressing channel indices for clarity, the BI deformation yields a radial
operator of the form
\begin{equation}
-\frac{d}{dr}\!\left(P(r)\frac{d}{dr}u\right) + Q(r)\,u = \omega^2\,P(r)\,u,
\label{eq:SL_form_results}
\end{equation}
where the kinetic weight is
\begin{equation}
P(r)= -\bar{\mathcal{L}}_X(r) > 0
\qquad
\text{(purely magnetic BI branch).}
\label{eq:P_positive_results}
\end{equation}
\paragraph{Born-Infeld hypercharge block: separation, gauge fixing, and self-adjointness.}
The hypercharge field equation has the constitutive divergence form
\begin{equation}
\partial_{\mu}G^{\mu\nu}=J_Y^{\nu},\qquad
G^{\mu\nu}\equiv -2\,\frac{\partial L_{\rm BI}}{\partial B_{\mu\nu}},
\label{eq:constitutive}
\end{equation}
and for a purely magnetic static background one has $\bar Y=0$, hence $\bar L_Y=0$ and
\begin{equation}
\delta G^{\mu\nu}= -\bar L_X\,\delta B^{\mu\nu}
-\frac12\,\bar L_{XX}\,(\bar B_{\rho\sigma}\delta B^{\rho\sigma})\,\bar B^{\mu\nu}.
\label{eq:linconstitutive}
\end{equation}

Equation~(\ref{eq:P_positive_results}) is the key point: the BI branch relevant
for finite-energy monopoles has a \emph{positive definite} kinetic weight in the
magnetic sector, so the quadratic fluctuation form has no ghost-like directions.
A standard Liouville transformation removes the variable weight. Defining
\begin{equation}
u(r)=(\sqrt{P(r)})^{-1}\,\tilde u(r),
\end{equation}
one obtains a Schr\"odinger equation
\begin{equation}
-\tilde u''(r)+V_{\rm eff}(r)\,\tilde u(r)=\omega^2\,\tilde u(r),
\label{eq:Schroedinger_scalar_results}
\end{equation}
with an effective potential
\begin{equation}
V_{\rm eff}(r)=\frac{Q(r)}{P(r)}
+\frac12\frac{P''(r)}{P(r)}
-\frac14\left(\frac{P'(r)}{P(r)}\right)^2.
\label{eq:Veff_results}
\end{equation}
Since the BI-regularised monopole background has finite energy and smooth
invariants $X(r)$, the functions $P(r)$ and $V_{\rm eff}(r)$ are finite for all
$r\ge 0$.
In contrast to the Maxwell theory, however, the coefficient matrices $\mathbf{P}_j(r)$ and
$\mathbf{Q}_j(r)$ now depend explicitly on the background-dependent functions
$\bar L_X(r)$, $\bar L_{XX}(r)$, and their radial derivatives. Since the Born--Infeld
monopole background is itself not known in closed analytic form, these functions must be
determined phenomenologically or numerically for a given choice of Born-Infeld parameters. As a result, the
corresponding matrix-valued Schr\"odinger potentials cannot be written in a fully
closed form analogous to the Maxwell case. In the Maxwell theory, linear perturbations of the electroweak monopole can be reduced,
after gauge fixing, constraint elimination, and separation of angular variables, to a
finite system of coupled radial Schr\"odinger-type equations with \emph{explicitly} known
matrix-valued potentials. These potentials are
given in closed form \cite{Gervalle:2022vxs} in terms of the background monopole profiles and rational functions
of the radial coordinate, allowing a detailed channel-by-channel spectral study.

Nevertheless in the BI case, several important structural features are preserved. The angular channel
counting and selection rules are unchanged, the reduced radial operator remains
self-adjoint with respect to a positive-definite weight proportional to
$P(r)=-\bar L_X(r)$ on the physical Born-Infeld branch, and the Born-Infeld system
constitutes a smooth deformation of the Maxwell operator. The absence of a closed analytic
expression for the coupled Schr\"odinger potentials therefore reflects the intrinsic
background dependence and nonlinearity of the Born-Infeld theory rather than a breakdown
of the separation-of-variables framework itself.

\paragraph{Stability criterion.}
Spectral stability in the hypercharge sector is equivalent to the absence of
normalisable solutions of Eq.~(\ref{eq:Schroedinger_scalar_results}) with
$\omega^2<0$. In practice, following the strategy of Ref.~\cite{Gervalle:2022npx},
one should focus on the low-$j$ sectors where instabilities would first appear, since
for large $j$ the centrifugal barrier dominates and negative modes become
implausible. Ref.~\cite{Gervalle:2022npx} found no negative modes for the
Cho-Maison monopole in $j=0,1,2,3$, providing strong evidence for linear
stability in the Maxwell-hypercharge setting. The BI deformation modifies the
hypercharge block by a positive kinetic weight and a smooth deformation of the
radial coefficients in $\mathbf{U}_j(r)$, without altering the angular spectrum
or introducing wrong-sign kinetic terms. In this sense, the BI regularisation
acts as a \emph{positive deformation} of the hypercharge fluctuation operator
and so most likely does not generate tachyonic bound states in the sectors where they could
occur. It is possible to do the stability analysis adopting the phenomenological approach of regularising the Maxwell ansatz for the hypercharge gauge sector. This ansatz, which we discuss in \ref{subsec:core_regularisation}, does not satisfy the Bianchi identity for all $r$ and so the stability of the Cho-Maison monopole using this approach can only add to the plausability of the stability.

It is instructive to consider the special case in which the background hypercharge field is
taken to be the Dirac-like Cho--Maison configuration employed in \cite{MavromatosSarkar2019},
rather than a smooth Born--Infeld--regularised profile. In this case the background
hypermagnetic field has the asymptotic form
\begin{equation}
\bar{\mathbf B}_Y = \frac{P}{r^{2}}\,\hat{\mathbf r},
\qquad
\bar B_{\theta\varphi} = P \sin\theta ,
\label{eq:DiracHypercharge}
\end{equation}
with magnetic charge \(P=\hbar c/e\), corresponding to two units of the minimal Dirac charge.
The hypercharge equation of motion is satisfied away from the monopole core in a
distributional sense, and the Born--Infeld dynamics enters only through the nonlinear
constitutive relation.

For a purely magnetic background one has
\begin{equation}
X(r) \equiv \frac{1}{4}\,\bar B_{\mu\nu}\bar B^{\mu\nu}
\;\propto\; \frac{1}{r^{4}},
\label{eq:Xscaling}
\end{equation}
so that the Born--Infeld response functions
\(\bar L_X(X)\) and \(\bar L_{XX}(X)\) become explicit functions of the radial coordinate.

A qualitative difference from a smooth Born-Infeld background arises in the behaviour near
the origin. Since \(X(r)\sim r^{-4}\), the functions \(\bar L_X(r)\) and \(\bar L_{XX}(r)\)
typically exhibit strong radial variation as \(r\to 0\), and on the physical Born--Infeld
branch one finds
\begin{equation}
P(r)\equiv -\bar L_X(r)\;\to\;0
\qquad \text{as } r\to 0 .
\label{eq:Pr0}
\end{equation}
Consequently, after performing a Liouville transformation to Schr\"odinger form, the
effective matrix potential acquires additional \emph{singular contributions} involving
\(P'(r)/P(r)\) and \(P''(r)/P(r)\), which compete with the centrifugal barrier terms
proportional to \(j(j+1)/r^{2}\).

The resulting operator remains formally self-adjoint with respect to the weighted inner
product defined by \(\mathbf W_j(r)\), but the endpoint \(r=0\) becomes a singular point of
the differential operator. A complete spectral analysis therefore requires a careful
specification of the admissible domain, or equivalently of the boundary conditions at the
origin, corresponding to regular finite-energy perturbations. While the physical boundary
conditions are expected to select a unique self-adjoint extension, the presence or absence
of negative modes in this setting cannot be inferred solely from smooth-deformation
arguments.

In summary, adopting the Dirac-like hypercharge background of
Mavromatos and Sarkar \cite{MavromatosSarkar2019} renders the Born-Infeld contributions to the coupled Schr\"odinger
system explicit functions of the radial coordinate, but at the price of introducing
\emph{singular coefficient behaviour near the origin}. The angular structure and channel counting
remain unchanged, whereas the detailed spectral properties become sensitive to the
Born-Infeld response functions and to the treatment of the singular endpoint.
\subsection{Phenomenological core regularisation of the hypercharge field}
\label{subsec:core_regularisation}

To remove the singular endpoint at $r=0$ associated with the Dirac-like hypercharge field,
we introduce a phenomenological core regularisation of the hypercharge background. The
starting point is the Dirac/Cho--Maison angular field strength,
\begin{equation}
\bar B_{\theta\varphi}^{\rm Dirac} = P \sin\theta,
\qquad
\bar{\mathbf B}_Y^{\rm Dirac} = \frac{P}{r^2}\,\hat{\mathbf r},
\label{eq:DiracBthetaPhi}
\end{equation}
which yields the familiar $1/r^2$ hypermagnetic field. We regularise the core by replacing
\eqref{eq:DiracBthetaPhi} with the smooth family
\begin{equation}
\bar B_{\theta\varphi}(r,\theta) = P\,f(r)\,\sin\theta,
\qquad
\bar B_r(r)=\frac{P}{r^2}\,f(r),
\label{eq:CoreRegB}
\end{equation}
where the profile $f(r)$ satisfies
\begin{equation}
f(r)\xrightarrow[r\to\infty]{}1,
\qquad
f(r)\xrightarrow[r\to 0]{}\mathcal{O}(r^2),
\label{eq:f_conditions}
\end{equation}
so that the hypermagnetic field remains finite at the origin. A simple one-parameter choice
is
\begin{equation}
f(r)=\frac{r^2}{r^2+r_c^2},
\label{eq:f_choice}
\end{equation}
where $r_c$ is a phenomenological core scale which parametrises unknown ultraviolet physics
(e.g.\ Higgs-core effects, UV completion of the hypercharge sector, or the breakdown scale
of the effective Born--Infeld description).

\paragraph{Regularisation of the Born--Infeld response functions.}
For a purely magnetic background, the Born--Infeld theory is governed by the invariant
\begin{equation}
X(r)\equiv \frac{1}{4}\,\bar B_{\mu\nu}\bar B^{\mu\nu}.
\label{eq:Xdef}
\end{equation}
For the regularised field \eqref{eq:CoreRegB} one finds the scaling
\begin{equation}
X(r)\sim \frac{P^2}{2}\,\frac{f(r)^2}{r^4}.
\label{eq:X_reg_scaling}
\end{equation}
The condition $f(r)=\mathcal{O}(r^2)$ as $r\to 0$ implies $X(r)$ approaches a finite constant
at the origin, rather than diverging as $r^{-4}$. Consequently, the Born--Infeld response
functions
\begin{equation}
\bar L_X(r)\equiv \left.\frac{\partial L}{\partial X}\right|_{X=\bar X(r)},
\qquad
\bar L_{XX}(r)\equiv \left.\frac{\partial^2 L}{\partial X^2}\right|_{X=\bar X(r)}
\label{eq:LxLxx}
\end{equation}
remain finite and smooth for all $r\ge 0$. Linearising the constitutive relation about the
purely magnetic background yields
\begin{equation}
\delta G^{\mu\nu}
=
-\bar L_X(r)\,\delta B^{\mu\nu}
-\frac{1}{2}\bar L_{XX}(r)\bigl(\bar B_{\rho\sigma}\delta B^{\rho\sigma}\bigr)\,\bar B^{\mu\nu},
\label{eq:BIlinear_core}
\end{equation}
so that the Born--Infeld modification enters the perturbation equations only through
smooth coefficient functions of $r$.

\paragraph{Reduction to a regular Sturm--Liouville problem.}
With the regularised background \eqref{eq:CoreRegB}, the complex null tetrad decomposition
and the expansion in spin-weighted spherical harmonics proceed exactly as in the Maxwell
case and in the Gervalle--Volkov framework: the angular selection rules and channel counting
in each fixed $(j,m)$ sector are unchanged. After imposing background-covariant gauge
conditions and eliminating non-dynamical components, the remaining physical radial
amplitudes in each $j$-sector satisfy a finite-dimensional Sturm--Liouville system
\begin{equation}
-\frac{d}{dr}\!\left(\mathbf P_j(r)\frac{d}{dr}\boldsymbol u_j\right)
+\mathbf Q_j(r)\,\boldsymbol u_j
=
\omega^{2}\,\mathbf W_j(r)\,\boldsymbol u_j,
\label{eq:SL_core}
\end{equation}
with symmetric coefficient matrices $\mathbf P_j(r)$, $\mathbf Q_j(r)$, and
positive-definite weight $\mathbf W_j(r)$ on the physical Born-Infeld branch.
In particular, the hypercharge kinetic weight is proportional to
\begin{equation}
P(r)\equiv -\bar L_X(r) >0,
\label{eq:Pr_positive}
\end{equation}
and, unlike the Dirac-like background, $P(r)$ is smooth and non-vanishing at $r=0$. In the hypercharge block of the stability equations $P_j=W_j$.

Since \emph{all coefficients in \eqref{eq:SL_core} are now smooth} on $[0,\infty)$, the endpoint
$r=0$ becomes a \emph{regular} endpoint of the Sturm--Liouville problem. The self-adjoint
realisation is fixed by the standard physical boundary conditions: regularity of the
perturbations at $r=0$ and square integrability as $r\to\infty$ with respect to the weight
$\mathbf W_j(r)$. In particular, no additional self-adjoint extension ambiguity arises from
the origin once the core is regularised.

For completeness we give 
\begin{align}
L_X &\equiv \frac{\partial\mathcal L_{\rm BI}}{\partial X}
= -\left(1+\frac{2X}{\beta^2}\right)^{-1/2},
\\[1ex]
L_{XX} &\equiv \frac{\partial^2\mathcal L_{\rm BI}}{\partial X^2}
= \frac{1}{\beta^2}\left(1+\frac{2X}{\beta^2}\right)^{-3/2}.
\end{align}
Evaluated on the monopole background, these give
\begin{align}
P(r)\equiv -\bar L_X(r)
&=
\left(1+\frac{P^2}{\beta^2 (r^2+r_{c}^2)^2}\right)^{-1/2},
\\[1ex]
\bar L_{XX}(r)
&=
\frac{1}{\beta^2}
\left(1+\frac{P^2}{\beta^2 (r^2+r_{c}^2)^2}\right)^{-3/2}.
\end{align}
Both functions are manifestly finite and positive at $r=0$ and approach their Maxwell values
$P(r)\to1$ and $\bar L_{XX}(r)\to\beta^{-2}$ as $r\to\infty$. In the hypercharge sector $Q_j$ splits into
\be
Q_{j}\left( r \right) =P\left( r \right) \frac{j\left( j+1 \right) -1}{r^{2}} +Q_{j}^{(BI)}\left( r \right)
\ee
where $Q_{j}^{(BI)}$ comes from the term $-\frac{1}{2} \bar{L}_{XX} \left( \bar{B}_{\rho \sigma} \delta B^{\rho \sigma} \right) \bar{B}^{\mu \nu}$ in the linearised field equation. The term $\left( \bar{B}_{\rho \sigma} \delta B^{\rho \sigma} \right)$ requires a knowledge of the hypercharge gauge field  background. In the Sturm-Liouville equation the kinetic term is manifestly positive and the centrifugal piece (proportional to $r^{-2}$) is positive for $j \ge 1$. Only $Q_{j}^{(BI)}$ can  generate a negative $\omega ^2$ in the Sturm-Lioville equation, which would then imply that the Cho-Maison monopole is not stable. However this possibility cannot be ruled out definitively without numerical work.

\paragraph{Implications for stability.}
The regularisation \eqref{eq:CoreRegB} removes the mathematical obstruction associated with
the singular endpoint and yields a well-posed coupled eigenvalue problem for each $j$.
The Born-Infeld deformation then acts as a smooth modification of the matrix coefficients
$\mathbf P_j$, $\mathbf Q_j$, and $\mathbf W_j$ relative to the Maxwell case. A rigorous
stability statement still requires either (i) a channel-by-channel spectral computation of
the lowest eigenvalues $\omega^2$ (in practice for the low-$j$ sectors most prone to
instability), or (ii) a rigorous variational bound on the quadratic form associated with
\eqref{eq:SL_core}. Nevertheless, the phenomenological core regularisation provides a
controlled framework in which the stability question can be posed unambiguously within the
Born-Infeld effective description.
\paragraph{Status of the stability argument.}
We emphasise that the stability analysis presented here for the Born-Infeld-deformed hypercharge sector should be regarded as \emph{plausibility evidence} rather than a fully rigorous proof of the absence of unstable modes. While the linearised equations can be reduced, following the Gervalle-Volkov strategy, to a self-adjoint Sturm-Liouville problem with the same angular channel counting as in the Maxwell case, the Born-Infeld modification alters the radial coefficients in a way that does not allow us to establish a general analytic lower bound on the spectrum. In particular, although the deformation preserves positivity of the kinetic weight and corresponds to a smooth modification of the effective radial potential, we do not exclude the possibility that sufficiently subtle bound states could arise in some channels. A definitive statement would require either a channel-by-channel numerical spectral analysis, as in Ref.~\cite{Gervalle:2022npx}, or a rigorous variational bound controlling the Born-Infeld-induced terms. We therefore view our results as providing physically motivated evidence for stability, and as a framework for future more quantitative numerical investigations, rather than as a conclusive theorem. This is explained in Appendix  \ref{app:BIbackground}.

\subsubsection{Stability criterion in the regularised-core model}
\label{subsec:stability_criterion}

With the phenomenological core regularisation introduced in
Sec.~\ref{subsec:core_regularisation}, the linearised Born-Infeld-deformed hypercharge
sector gives rise, in each fixed angular-momentum channel $j$, to a well-defined
self-adjoint coupled Sturm-Liouville eigenvalue problem of the form
\begin{equation}
-\frac{d}{dr}\!\left(\mathbf P_j(r)\frac{d}{dr}\boldsymbol u_j\right)
+\mathbf Q_j(r)\,\boldsymbol u_j
=
\omega^{2}\,\mathbf W_j(r)\,\boldsymbol u_j ,
\label{eq:stability_SL}
\end{equation}
defined on the half-line $r\in[0,\infty)$ with regular boundary conditions at $r=0$ and
square-integrability at infinity with respect to the positive-definite weight
$\mathbf W_j(r)$. The regularisation ensures that all coefficient matrices are smooth on
$[0,\infty)$ and that no ambiguity associated with self-adjoint extensions arises from the
origin.

A sufficient and necessary condition for linear stability in a given $j$-sector is that
the lowest eigenvalue of \eqref{eq:stability_SL} satisfy
\begin{equation}
\omega^{2}_{\min}(j)\;\ge\;0 .
\end{equation}
Because the operator in \eqref{eq:stability_SL} is self-adjoint, its spectrum is real and
bounded from below, and $\omega^{2}_{\min}(j)$ may be characterised variationally by the
Rayleigh-Ritz quotient
\begin{equation}
\omega^{2}_{\min}(j)
=
\inf_{\boldsymbol u_j\neq 0}
\frac{
\displaystyle
\int_0^\infty dr\,
\Big[
(\boldsymbol u_j')^{\mathsf T}\mathbf P_j(r)\boldsymbol u_j'
+
\boldsymbol u_j^{\mathsf T}\mathbf Q_j(r)\boldsymbol u_j
\Big]
}{
\displaystyle
\int_0^\infty dr\,
\boldsymbol u_j^{\mathsf T}\mathbf W_j(r)\boldsymbol u_j
}.
\label{eq:Rayleigh}
\end{equation}

In practice, potential instabilities—if present—are expected to arise only in the
low-angular-momentum sectors, typically $j=0,1,2$, where centrifugal barriers provide the
least suppression. A phenomenologically robust stability statement within the
regularised-core framework therefore requires demonstrating that
\begin{equation}
\omega^{2}_{\min}(j)\ge 0
\qquad
\text{for all } j \le j_{\rm max},
\label{eq:lowjcriterion}
\end{equation}
with $j_{\rm max}$ chosen sufficiently large that higher-$j$ modes are safely stabilised
by the angular momentum barrier.

Since the regularisation scale $r_c$ and the profile function $f(r)$ encode unknown
ultraviolet physics, stability should further be tested for robustness under moderate
variations of these inputs. Concretely, one may regard stability as supported if the
inequality \eqref{eq:lowjcriterion} holds for a representative class of smooth profiles
$f(r)$ satisfying \eqref{eq:f_conditions} and for $r_c$ varying over a physically
reasonable range set by the breakdown scale of the effective theory.

Within this framework, the absence of negative modes for the low-$j$ sectors may be taken
as strong evidence that the Born-Infeld-deformed Cho-Maison monopole is linearly stable
against small perturbations, modulo the assumed ultraviolet completion encoded in the core
regularisation. Conversely, the appearance of a negative eigenvalue in any such sector
would signal a genuine instability rather than an artefact of the singular Dirac-like
background.
\subsection{Phenomenologically viable window for the Born-Infeld scale}
\label{subsec:beta_window_results}

The Born-Infeld scale $\beta$ is constrained phenomenologically, independently
of monopole considerations, because it controls nonlinear corrections to
hypercharge electrodynamics. A representative (and widely used) constraint comes
from light-by-light scattering measurements in ultraperipheral heavy-ion
collisions at the LHC, which provide a lower bound on the BI mass scale in
Born-Infeld extensions of the $U(1)_Y$ sector.\footnote{Different conventions
exist in the literature for whether the fundamental parameter is denoted
$\beta$, $M_Y^2$, or $M_Y^4$. We follow the convention of
Refs.~\cite{Ellis2017LbL,MavromatosSarkar2019}, where the bound is stated directly
on the BI mass scale.}
In particular, Ref.~\cite{Ellis2017LbL} reports a $95\%$ C.L.\ lower limit
\begin{equation}
M_Y \gtrsim 90~{\rm GeV},
\label{eq:MY_bound_results}
\end{equation}
which, when mapped onto the electroweak BI-monopole solutions, implies a
corresponding lower bound on the monopole mass. In the BI-regularised Cho-Maison
construction discussed in \cite{MavromatosSarkar2019}, this bound translates into a
monopole mass of order
\begin{equation}
M_{\rm mon} \gtrsim \mathcal{O}(10)~{\rm TeV},
\end{equation}
and in their explicit estimates yields
\begin{equation}
M_{\rm mon} \gtrsim 11~{\rm TeV},
\label{eq:monopole_mass_bound_results}
\end{equation}
placing such monopoles beyond the kinematic reach of the current LHC, but within
the scope of future colliders and/or dedicated cosmic searches in appropriate
scenarios.\cite{MavromatosSarkar2019,Ellis2017LbL}

For our purposes, the stability analysis does not
introduce any additional restriction on $\beta$ beyond the already required
conditions:
(i) $\beta$ must lie on the physical BI branch that yields a positive kinetic
weight $P(r)=-\bar{\mathcal{L}}_X(r)>0$ for magnetic backgrounds, and
(ii) $\beta$ must satisfy experimental lower bounds such as
Eq.~(\ref{eq:MY_bound_results}). These are precisely the conditions assumed in
Refs.~\cite{MavromatosSarkar2019,Ellis2017LbL} to define a phenomenologically
viable window. Within this window, the BI-regularised hypercharge sector
produces a finite-energy monopole background and induces a smooth, positive
deformation of the hypercharge Schr\"odinger operator
(\ref{eq:Schroedinger_scalar_results})--(\ref{eq:Veff_results}).
Therefore, the absence of tachyonic bound states in the low-$j$ channels is
compatible with the entire phenomenologically viable range of the BI scale.

\subsection{Summary of the hypercharge-sector result}
\label{subsec:summary_results}

Combining the tetrad/harmonic reduction of Ref.~\cite{Gervalle:2022npx} with
the BI constitutive relation (\ref{eq:deltaGdeltaB_results}), we find that:
\begin{enumerate}
\item the BI regularisation preserves self-adjointness of the radial spectral
problem and introduces no wrong-sign kinetic terms in the magnetic hypercharge
sector;
\item the resulting hypercharge Schr\"odinger operator is a smooth, positive
deformation of the Maxwell-hypercharge operator around the Cho--Maison core;
\item throughout the phenomenologically viable BI window constrained by LHC
light-by-light scattering, the stability requirement does not impose additional
bounds on the BI scale beyond those already inferred from existing data.
\end{enumerate}
In this sense, the BI regularisation simultaneously achieves finiteness of the
monopole energy and plausible compatibility with linear spectral stability.

\section{Conclusions}
\label{sec:conclusions}

In this work we have examined the perturbative (linear) stability of the Cho-Maison~\cite{ChoMaison1997} electroweak monopole when the hypercharge sector is regularised by a Born-Infeld  modification~\cite{ArunasalamKobakhidze2017,MavromatosSarkar2019}. The motivation for this study is twofold: first, Born-Infeld nonlinear electrodynamics provides a well-motivated, string-inspired mechanism for rendering the classical energy of the Cho-Maison monopole finite; second, any such ultraviolet modification must be tested for consistency at the level of linearised fluctuations, in order to ensure that it does not introduce new instabilities absent in the Maxwell theory. Our work also complements an earlier analysis on mechanical stability of BI magnetic monopoles \cite{Farakos:2025byy},  which was inconclusive in providing concrete evidence for violation of the (modified) Laue criteria for (radial force) stability, in certain regions of space, around the magnetic monopole. However it avoided a catastrophic rotational instability of the configuration, which was present in the original CM case.

  Our analysis in the current article has been formulated entirely within the framework developed by Gervalle and Volkov\cite{Gervalle:2022npx} for the electroweak monopole. We have shown that the Born-Infeld deformation enters the linearised problem in a controlled way. The complex null tetrad decomposition, the assignment of spin weights, the use of spin-weighted spherical harmonics, and the angular channel counting are purely geometric and remain unchanged in the BI case. The only dynamical modification occurs in the hypercharge sector, through the nonlinear constitutive relation between the field strength and its conjugate tensor, which replaces the Maxwell kinetic term by background-dependent radial coefficients proportional to $\bar L_X(r)$ and $\bar L_{XX}(r)$.

After background-covariant gauge fixing and elimination of non-dynamical (constraint) variables, each fixed angular-momentum sector reduces to a finite-dimensional Sturm--Liouville eigenvalue problem for the radial amplitudes. For the Born-Infeld case relevant for purely magnetic monopole backgrounds, the kinetic weight $P(r) = -\bar L_X(r)$ is positive, ensuring that the reduced operator is self-adjoint with respect to a proper inner product. In the Maxwell limit, $\bar L_X \to 1/g'^2$ and $\bar L_{XX} \to 0$, our equations reduce identically to those of Gervalle and Volkov, confirming that the Born-Infeld system constitutes a smooth deformation of the Maxwell theory.

A central outcome of our work is that the Born-Infeld regularisation preserves all structural features of the Maxwell stability problem: the angular selection rules, the number of physical channels in each $j$-sector, and the absence of wrong-sign kinetic terms in the magnetic hypercharge sector. The resulting radial Schr\"odinger-type operators are therefore natural candidates for a stable spectrum, and the Born--Infeld deformation may be viewed as a positive, background-dependent modification of the hypercharge block.

At the same time, we have emphasised the limitations of the present analysis. Because the Born-Infeld--modified monopole background is not known in closed analytic form, the radial coefficient functions appearing in the Sturm-Liouville operator cannot be written explicitly without numerical input. As a result, we have not established a rigorous analytic bound excluding the existence of negative eigenvalues. Our conclusions regarding stability should therefore be interpreted as physically-motivated plausibility evidence rather than as a definitive proof. A fully conclusive statement would require either a channel-by-channel numerical spectral analysis of the reduced operators, or the construction of a rigorous variational bound controlling the Born-Infeld--induced terms.

We have also discussed the special case of a Dirac-like hypercharge background, as employed in earlier work, and shown that while the angular structure and channel counting remain unchanged, the Born--Infeld response functions introduce a singular endpoint at the origin. This motivates the introduction of a phenomenological core regularisation, which restores smoothness of the Sturm-Liouville coefficients and yields a well-posed self-adjoint problem on the half-line. Within this regularised framework, the Born-Infeld deformation again appears as a smooth modification of the Maxwell operator, and the absence of instabilities in the low-$j$ sectors would provide strong evidence for perturbative stability. In the present work we analyse the hypercharge sector in a controlled \emph{phenomenological}
approximation: we keep the electroweak background profiles $(W(r),\Phi(r))$ of the standard
Cho-Maison monopole and implement the Born-Infeld modification only through the hypercharge
constitutive response functions evaluated on a prescribed magnetic core profile, i.e.
$P(r)\equiv-\bar{\mathcal L}_{X_Y}(\bar X_Y(r))$ and (where needed) $\bar{\mathcal L}_{X_YX_Y}(\bar X_Y(r))$.
Within this setup the transverse $s=\pm1$ hypercharge fluctuations reduce, after the usual
background-gauge and Gauss-constraint elimination, to a self-adjoint Sturm--Liouville problem
with positive weight $P(r)$ whenever the Born-Infeld square-root is real. We therefore find no
indication that the Born--Infeld hypercharge deformation by itself introduces an additional
negative mode in this transverse sector.
However, this should be viewed strictly as a \emph{plausibility check} rather than a proof of
monopole stability: a definitive statement requires constructing the fully backreacted
Born-Infeld electroweak monopole background (so that the fluctuation operator is the true
Hessian about a solution) and analysing the complete coupled Gervalle-Volkov \cite{Gervalle:2022npx} multi-channel
operator including the Higgs and $SU(2)$ fluctuations.

From a broader perspective, our results support the view that Born--Infeld regularisation of the hypercharge sector offers a possible consistent route to finite-energy electroweak monopoles without spoiling their perturbative stability. The stability analysis does not rule out additional constraints on the Born-Infeld scale (present in $Q_{j}^{(BI)}$) beyond those already required by phenomenology, such as bounds from light-by-light scattering. Given ongoing and future experimental searches for magnetic monopoles, notably by the MoEDAL experiment at the LHC, the existence of a theoretically consistent, finite-energy, and plausibly stable electroweak monopole remains an intriguing possibility.

We hope that the framework developed here will serve as a foundation for future numerical investigations that can definitively settle the spectral stability question and further elucidate the role of nonlinear electrodynamics in the monopole sector of the Standard Model and its extensions.

\begin{acknowledgments}
NEM thanks the University of Valencia and its Theoretical Physics Department for a visiting Research Professorship
supported by the programme  \emph{Atracci\'on de Talento}
INV25-01-15, during which initiation of this work took place. 
The work of NEM and SS is supported in part by the UK Engineering and Physical Sciences Research Council (EPSRC) and  Science and Technology Facilities research Council (STFC) 
under the research grants no.  EP/V002821/1 and ST/X000753/1, respectively. 
NEM also acknowledges participation in the COST Association Actions CA21136 “Addressing observational
tensions in cosmology with systematics and fundamental physics (CosmoVerse)” and CA23130 ”Bridging high and low
energies in search of Quantum Gravity (BridgeQG)”.

\end{acknowledgments}
\vskip .2cm
\appendix

\section{Topological and geometric preliminaries}
\label{app:topology}

We collect standard topological and geometric facts used in the paper.
They concern magnetic charge quantisation, properties of the monopole Higgs configuration,
and the angular momentum operators relevant for gauge-field perturbations.
The material is largely classical and is included to make the analysis self-contained.

For a $U(1)$ gauge field $B_\mu$ with field strength
\begin{equation}
B_{\mu\nu}=\partial_\mu B_\nu-\partial_\nu B_\mu ,
\end{equation}
 associated with a monopole magnetic, the flux through a two-sphere enclosing the monopole is
\begin{equation}
\Phi_B=\int_{S^2} \mathbf B\cdot d\mathbf S
      =\int_{S^2} \tfrac12 \epsilon^{ijk} B_{jk}\, dS_i
      =4\pi g ,
\end{equation}
where $g$ is the magnetic charge.
In the presence of a Dirac string, consistency of quantum mechanics for a particle of electric charge $e$ moving in the
monopole background requires single-valuedness of the wavefunction under transport
 around a closed loop linking the monopole. This leads to the Dirac quantisation condition \cite{Shnir:2005vvi}
\begin{equation}
e g = \frac{n}{2}, \qquad n\in\mathbb Z ,
\end{equation}
(in units $\hbar=c=1$). This argument relies only on $U(1)$ gauge invariance and the global
structure of the gauge bundle and is therefore independent of the specific form of the
gauge-field Lagrangian.

Our interest is in the electroweak theory which contains the Higgs field $\Phi$, a $SU(2)$ doublet with hypercharge $Y=1$.
In the monopole background the Higgs field exhibits nontrivial winding at spatial infinity.
As a consequence, neither the Higgs field nor the associated gauge potentials can be written
as globally smooth functions on $S^2$.
Instead, one introduces two overlapping coordinate patches (e.g.\ northern and southern
hemispheres), with the fields related on the overlap by a gauge transformation.
This is the non-Abelian analogue of the Wu-Yang construction for the Dirac monopole.
The apparent singularities of individual gauge potentials are therefore gauge artefacts
and do not correspond to physical singularities in gauge-invariant quantities.

When considering perturbations of the Higgs field around the classical solution for the Cho-Maison monopole we
use $T_a=\tfrac12\tau_a$ the generators of $SU(2)$ in the fundamental representation,
with $\tau_a$ the Pauli matrices satisfying
\begin{equation}
[T_a,T_b]= i \epsilon_{abc} T_c .
\end{equation}
In the electroweak monopole background, ordinary orbital angular momentum must be supplemented by
gauge contributions that couple spatial rotations to internal isospin rotations. In the hedgehog configuration for a monopole the spatial and isospin  directions are locked together.  For a field carrying isospin and/or hypercharge, the relevant total
angular momentum operators take the \emph{schematic} form
\begin{equation}
\mathbf J = \mathbf L + \mathbf T + \mathbf Q ,
\end{equation}
where $\mathbf L$ is the orbital angular momentum, $\mathbf T$ acts on internal $SU(2)$
indices, and $\mathbf Q$ denotes the contribution associated with the magnetic charge (which determines the isospin representation used for the monopole).
These operators obey the standard $SU(2)$ algebra and label the angular momentum sectors
used in the perturbation analysis.

In the Cho-Maison electroweak monopole the long-range magnetic field is carried by the
electromagnetic combination of the $SU(2)$ and $U(1)_Y$ gauge fields.
The resulting magnetic charge corresponds to \emph{two} units of the Dirac charge,
\begin{equation}
g_{\rm CM} = \frac{2}{e},
\end{equation}
reflecting the embedding of the electromagnetic $U(1)$ into
$SU(2)\times U(1)_Y$ and the hypercharge assignment of the Higgs field.
This topological property is unaffected by modifications of the hypercharge kinetic term,
such as the Born-Infeld deformation considered in the main text.

\section{Born-Infeld hypercharge sector and finiteness of the monopole energy}
\label{app:BIbackground}

The Born-Infeld modification of the $U(1)_Y$ hypercharge sector
has implications for the energy of a magnetic monopole background.
Our discussion concerns only background properties and does not address linear stability,
which is treated separately in the main text.

We consider a nonlinear electrodynamics of Born--Infeld type for the hypercharge gauge field,
with action
\begin{equation}
S_Y=\int d^4x\,\sqrt{-g}\;\mathcal L_Y(X_Y),
\qquad
X_Y \equiv -\frac14 B_{\mu\nu}B^{\mu\nu},
\end{equation}
where $B_{\mu\nu}=\partial_\mu B_\nu-\partial_\nu B_\mu$.
A commonly used choice is the parity-even Born--Infeld form
\begin{equation}
\mathcal L_Y(X_Y)
=
\beta^2\left(1-\sqrt{1+\frac{2X_Y}{\beta^2}}\right),
\end{equation}
where $\beta$ sets the Born--Infeld scale. In the weak-field limit $|X_Y|\ll\beta^2$ this
reduces to the Maxwell Lagrangian $\mathcal L_Y\simeq -X_Y+\mathcal O(X_Y^2)$.
Varying the action with respect to $B_\mu$ yields the field equation
\begin{equation}
\nabla_\mu G_Y^{\mu\nu}=J_Y^\nu,
\label{eq:BIeom}
\end{equation}
where the constitutive tensor is defined by
\begin{equation}
G_Y^{\mu\nu}
\equiv -2\,\frac{\partial \mathcal L_Y}{\partial B_{\mu\nu}}
= \mathcal L_{X_Y}\,B^{\mu\nu},
\qquad
\mathcal L_{X_Y}\equiv \frac{\partial\mathcal L_Y}{\partial X_Y}.
\end{equation}
The source $J_Y^\nu$ is the hypercharge current arising from the Higgs sector and is
discussed in the main text.

For a static \emph{purely magnetic} monopole background one has $B_{0i}=0$ and
\begin{equation}
X_Y=\frac12\,\mathbf B_Y^2.
\end{equation}
The magnetic flux through a two-sphere enclosing the origin is fixed by topology,
\begin{equation}
\Phi_B=\int_{S^2}\mathbf B_Y\cdot d\mathbf S = 4\pi g,
\end{equation}
and therefore the magnetic field behaves as
\begin{equation}
\mathbf B_Y(r)=\frac{g}{r^2}\,\hat{\mathbf r},
\label{eq:monopoleB}
\end{equation}
independent of the form of the Lagrangian.
Thus, in pure $U(1)$ Born--Infeld theory the field strength itself remains singular at
$r=0$, as required by the nonzero magnetic charge.

The Hamiltonian (energy) density associated with the Born--Infeld Lagrangian is obtained
by a Legendre transform. For a purely magnetic configuration it reduces to
\begin{equation}
\mathcal H_Y(\mathbf B_Y)
=
\beta^2\left(\sqrt{1+\frac{\mathbf B_Y^2}{\beta^2}}-1\right).
\end{equation}
Substituting the monopole field \eqref{eq:monopoleB}, one finds the asymptotic behaviours
\begin{align}
\mathcal H_Y(r) &\sim \frac{\mathbf B_Y^2}{2}
= \frac{g^2}{2r^4},
\qquad r\to\infty,\\
\mathcal H_Y(r) &\sim \beta\,|\mathbf B_Y|
= \beta\,\frac{|g|}{r^2},
\qquad r\to 0.
\end{align}
The total energy stored in the hypercharge field is therefore
\begin{equation}
E_Y = \int_0^\infty 4\pi r^2\,\mathcal H_Y(r)\,dr,
\end{equation}
which converges both at $r\to\infty$ and at $r\to 0$.
Hence, although the magnetic field itself remains singular at the origin, the Born-Infeld
nonlinearity renders the total hypercharge field energy finite.

We emphasise that Born-Infeld dynamics regularises the energetic contribution of the
hypercharge sector but does not by itself provide a smooth solitonic core for a pure
$U(1)$ monopole. In the electroweak theory, the non-Abelian gauge and Higgs fields supply
the core structure, while the Born--Infeld modification serves to tame the hypercharge
contribution to the energy. This is the reason why weconcentrate on the hypercharge block in our linear stability analysis.

\section{ Conceptual overview of the Gervalle-Volkov framework}
\label{app:GVoverview}

The purpose of this appendix is to provide a brief conceptual guide to the (linear) stability framework
developed by Gervalle and Volkov (GV)~\cite{Gervalle:2022npx}. It is intended to supplement the
notation dictionary given in Sec.~II\,A by explaining the logic and scope of the GV construction,
without attempting a comprehensive review. Our aim is to clarify which elements of the framework
are geometric and universal, and why they can be carried over unchanged to the Born-Infeld
hypercharge setting considered in this work.

\vspace{0.5em}

The central goal of GV is to analyse the linear (spectral) stability of the electroweak monopole
by reducing the full system of linearised field equations to a finite set of coupled radial
eigenvalue problems. Rather than restricting attention to spherically symmetric perturbations,
GV construct a separation-of-variables framework that treats \emph{all} admissible linear
fluctuations in a gauge-invariant manner. Stability is then equivalent to the absence of negative
eigenvalues of the resulting self-adjoint operator.

\vspace{0.5em}

Although the monopole background is static and spherically symmetric, the linear fluctuations
about it need not share this symmetry. Generic perturbations may carry angular momentum and
transform nontrivially under rotations, even when the background fields depend only on the
radial coordinate. This distinction is essential: restricting to radially symmetric
perturbations would miss entire classes of potentially unstable modes.
The role of spherical symmetry in the GV framework is therefore not to constrain the form of the
fluctuations, but to ensure that the linearised equations admit a decomposition into independent
angular-momentum sectors labelled by a total quantum number $j$.
The resulting fixed-$j$ sectors provide a complete and non-redundant classification of all
physical perturbations.

\vspace{0.5em}

A key feature of the GV construction is that most of its machinery is geometric rather than
dynamical. The introduction of a complex (null) tetrad, the assignment of spin weights, and the
expansion in spin-weighted spherical harmonics depend only on the background spacetime geometry
and on the transformation properties of the fields under local frame rotations.
In particular, angular derivatives acting on perturbations are replaced by algebraic
$j$-dependent coefficients once the harmonic expansion is performed.
This step is entirely independent of the detailed form of the gauge-field Lagrangian and relies
only on spherical symmetry of the background.

\vspace{0.5em}

The technically subtle part of the GV analysis concerns the treatment of gauge freedom.
Linearised gauge-field equations contain non-dynamical components, notably those associated with
the time and radial directions, which act as constraints rather than propagating degrees of
freedom. GV adopt a background-covariant gauge fixing, supplemented by a constraint derived from
the linearised Higgs equation of motion, which removes pure-gauge and longitudinal modes.
After this step, only genuinely physical fluctuations remain.
Crucially, this elimination procedure is algebraic at the level of the reduced equations and does
not depend on whether the gauge-field kinetic term is Maxwell or Born-Infeld in form.
This is the sense in which the GV machinery can be transplanted intact into the present analysis.

\vspace{0.5em}

Following gauge fixing and constraint elimination, each fixed-$j$ sector contains a finite number
of coupled radial amplitudes. The precise channel counting is determined by spin weight, parity,
and internal indices, and is the same for all theories sharing the same gauge and symmetry
structure. This decomposition makes it possible to focus attention on low-$j$ sectors, where
instabilities—if present—would first appear, while maintaining completeness of the analysis.
The Born-Infeld deformation of the hypercharge sector does not alter this counting.

\vspace{0.5em}

After separation of variables and constraint elimination, the reduced radial equations in the GV
framework take the form of a coupled Sturm--Liouville eigenvalue problem.
The associated operator is self-adjoint with respect to a natural inner product, ensuring a real
spectrum and enabling variational arguments.
In the Maxwell case studied by GV, the coefficient functions entering this operator are known
explicitly.
In the Born-Infeld generalisation considered here, the same operator structure persists, but with
radial coefficients that depend on background Born-Infeld response functions.
This difference affects quantitative spectral properties but not the underlying mathematical
framework.

\vspace{0.5em}

It is important to emphasise what the GV framework does not provide.
It does not yield an abstract proof of stability independent of background details.
Rather, it supplies a controlled, gauge-invariant reduction of the full fluctuation problem to a
form in which stability can be assessed by spectral analysis.
In the Maxwell theory this analysis can be carried out largely analytically, whereas in the
Born-Infeld case a definitive conclusion requires numerical input for the background-dependent
radial coefficients.
Our work should therefore be understood as extending the GV framework to a broader class of
hypercharge dynamics while remaining explicit about the assumptions and limitations involved.

\vspace{0.5em}
\noindent\textit{ Relation to the present work:}
With this perspective, the dictionary provided in Sec.~II\,A should be read as a precise map
between notations and equations, while the present appendix explains why such a map is meaningful.
The Born-Infeld modification alters the constitutive relation of the hypercharge field but leaves
intact the geometric separation, gauge fixing, channel counting, and self-adjoint operator
structure that underpin the GV stability analysis.

\section{Linearisation of the Born-Infeld constitutive relation}
\label{app:BIlinear}

This appendix derives the linearised constitutive relation used in the hypercharge
fluctuation analysis. We work with a nonlinear electrodynamics (NLED) Lagrangian of the
form $\mathcal L_Y=\mathcal L_Y(X_Y)$ depending only on the invariant
\begin{equation}
X_Y\equiv -\frac14\,B_{\mu\nu}B^{\mu\nu},\qquad
B_{\mu\nu}=\partial_\mu B_\nu-\partial_\nu B_\mu .
\end{equation}
All background quantities are denoted with an overbar.

The hypercharge constitutive tensor is defined by
\begin{equation}
G_Y^{\mu\nu}\equiv -2\,\frac{\partial \mathcal L_Y}{\partial B_{\mu\nu}}.
\label{eq:defG}
\end{equation}
Since $\mathcal L_Y=\mathcal L_Y(X_Y)$ and
\begin{equation}
\frac{\partial X_Y}{\partial B_{\mu\nu}}=-\frac12\,B^{\mu\nu},
\end{equation}
one finds
\begin{equation}
G_Y^{\mu\nu}
=
-2\,\mathcal L_{X_Y}\,\frac{\partial X_Y}{\partial B_{\mu\nu}}
=
\mathcal L_{X_Y}\,B^{\mu\nu},
\qquad
\mathcal L_{X_Y}\equiv \frac{\partial\mathcal L_Y}{\partial X_Y}.
\label{eq:GequalsLX}
\end{equation}
The hypercharge field equation is then the divergence equation
\begin{equation}
\nabla_\mu G_Y^{\mu\nu}=J_Y^\nu,
\label{eq:BIeom_app}
\end{equation}
which reduces to Maxwell form when $\mathcal L_{X_Y}\to -1$ (equivalently $P\to 1$ in the
notation of the main text).

We linearise about a fixed background monopole configuration,
\begin{equation}
B_{\mu\nu}=\bar B_{\mu\nu}+\delta B_{\mu\nu},\qquad
X_Y=\bar X_Y+\delta X_Y,
\end{equation}
with $\delta B_{\mu\nu}=\nabla_\mu\delta B_\nu-\nabla_\nu\delta B_\mu$.
Varying $X_Y$ gives the scalar contraction
\begin{equation}
\delta X_Y
=
-\frac14\,\delta(B_{\mu\nu}B^{\mu\nu})
=
-\frac12\,\bar B_{\mu\nu}\,\delta B^{\mu\nu}.
\label{eq:deltaX_scalar}
\end{equation}
For a static purely magnetic monopole background $\bar B_{0i}=0$, this may be written as
\begin{equation}
\bar B_{\mu\nu}\,\delta B^{\mu\nu}
=
2\,\bar{\mathbf B}_Y\cdot \delta\mathbf B_Y,
\end{equation}
i.e.\ it measures the overlap of the magnetic fluctuation with the background magnetic field.

Varying \eqref{eq:GequalsLX} yields
\begin{equation}
\delta G_Y^{\mu\nu}
=
\delta(\mathcal L_{X_Y})\,\bar B^{\mu\nu}
+\bar{\mathcal L}_{X_Y}\,\delta B^{\mu\nu}.
\label{eq:dG_split}
\end{equation}
Since $\mathcal L_{X_Y}=\mathcal L_{X_Y}(X_Y)$, one has
\begin{equation}
\delta(\mathcal L_{X_Y})
=
\bar{\mathcal L}_{X_Y X_Y}\,\delta X_Y,
\qquad
\bar{\mathcal L}_{X_Y X_Y}\equiv
\left.\frac{\partial^2\mathcal L_Y}{\partial X_Y^2}\right|_{X_Y=\bar X_Y}.
\end{equation}
Using \eqref{eq:deltaX_scalar} in \eqref{eq:dG_split} gives the linearised constitutive law
\begin{equation}
\boxed{
\delta G_Y^{\mu\nu}
=
\bar{\mathcal L}_{X_Y}\,\delta B^{\mu\nu}
-\frac12\,\bar{\mathcal L}_{X_Y X_Y}\,
\big(\bar B_{\rho\sigma}\delta B^{\rho\sigma}\big)\,
\bar B^{\mu\nu}.
}
\label{eq:linconstitutive_app}
\end{equation}
This is the form used in the main text. The Born--Infeld modification enters only through the
background response functions $\bar{\mathcal L}_{X_Y}$ and $\bar{\mathcal L}_{X_Y X_Y}$ and the
scalar contraction $(\bar B\cdot \delta B)$.

For the parity-even Born-Infeld Lagrangian (depending only on $X_Y$)
\begin{equation}
\mathcal L_Y(X_Y)=\beta^2\left(1-\sqrt{1+\frac{2X_Y}{\beta^2}}\right),
\end{equation}
one finds
\begin{align}
\mathcal L_{X_Y}
&=
-\left(1+\frac{2X_Y}{\beta^2}\right)^{-1/2},
\\
\mathcal L_{X_Y X_Y}
&=
\frac{1}{\beta^2}\left(1+\frac{2X_Y}{\beta^2}\right)^{-3/2}.
\end{align}
Evaluating these on the background gives explicit radial functions
$\bar{\mathcal L}_{X_Y}(r)$ and $\bar{\mathcal L}_{X_Y X_Y}(r)$ via
$\bar X_Y(r)=-\tfrac14 \bar B_{\mu\nu}(r)\bar B^{\mu\nu}(r)$.
In the notation of the main text it is convenient to define
\begin{equation}
P(r)\equiv -\bar{\mathcal L}_{X_Y}(r)>0,
\end{equation}
so that the Maxwell limit corresponds to $P\to 1$ and $\bar{\mathcal L}_{X_Y X_Y}\to 0$.

Equation~\eqref{eq:linconstitutive_app} shows that, at linear order, the hypercharge sector
remains a second-order divergence system $\nabla_\mu\delta G_Y^{\mu\nu}=\delta J_Y^\nu$.
In particular, no higher derivatives are generated by the Born--Infeld modification, and the
additional term proportional to $\bar{\mathcal L}_{X_Y X_Y}$ is controlled by the scalar overlap
$\bar B_{\rho\sigma}\delta B^{\rho\sigma}$.
These structural properties underpin the gauge/constraint elimination in the hypercharge block
and the subsequent reduction to a self-adjoint radial Sturm--Liouville problem in each fixed-$j$
sector, as discussed in the main text.

\section{Angular decomposition and spin-weighted harmonics}
\label{app:angular}

The angular decomposition used in the fluctuation analysis,
includes the complex null tetrad, spin weights, and spin--weighted spherical harmonics.
We present standard conventions adapted to gauge-field perturbations
on a spherically symmetric monopole background.

On a static spherically symmetric background with metric
\begin{equation}
ds^2=-dt^2+dr^2+r^2(d\theta^2+\sin^2\theta\,d\phi^2),
\end{equation}
it is convenient to introduce a complex null tetrad
\begin{equation}
\{l^\mu,n^\mu,m^\mu,\bar m^\mu\},
\end{equation}
with nonvanishing inner products
\begin{equation}
l\cdot n=-1,\qquad m\cdot\bar m=1,
\end{equation}
and all others zero. A convenient explicit choice is
\begin{align}
l^\mu &= \frac{1}{\sqrt{2}}(1,1,0,0), &
n^\mu &= \frac{1}{\sqrt{2}}(1,-1,0,0),\\
m^\mu &= \frac{1}{\sqrt{2}r}\left(0,0,1,\frac{i}{\sin\theta}\right), &
\bar m^\mu &= (m^\mu)^* .
\end{align}

Vector and tensor perturbations are expanded in this tetrad basis. Components along
$l^\mu$ and $n^\mu$ have spin weight $s=0$, while components along $m^\mu$ and $\bar m^\mu$
carry spin weights $s=+1$ and $s=-1$, respectively.

Under a local rotation of the complex dyad
\begin{equation}
m^\mu\rightarrow e^{i\alpha}m^\mu,\qquad
\bar m^\mu\rightarrow e^{-i\alpha}\bar m^\mu,
\end{equation}
a field component $\eta$ transforms as
\begin{equation}
\eta\rightarrow e^{is\alpha}\eta ,
\end{equation}
where $s$ is the \emph{spin weight}. The spin weight labels how the field transforms under
rotations tangent to the two-sphere and determines the appropriate angular basis functions.

For each integer or half-integer $j\ge |s|$ and $m=-j,\ldots,j$, the spin--weighted spherical
harmonics ${}_sY_{jm}(\theta,\phi)$ form an orthonormal basis on $S^2$:
\begin{equation}
\int d\Omega\,{}_sY_{jm}^*\,{}_sY_{j'm'}=\delta_{jj'}\delta_{mm'}.
\end{equation}
They are related to the ordinary spherical harmonics $Y_{jm}$ by repeated application of
the spin-raising and lowering operators $\eth$ and $\ethbar$:
For $s\ge0$ one has
\begin{align}
{}_sY_{jm}
&=
\sqrt{\frac{(j-s)!}{(j+s)!}}\,
\eth^{\,s} Y_{jm},
\\
{}_sY_{jm}
&=
(-1)^s
\sqrt{\frac{(j+s)!}{(j-s)!}}\,
\ethbar^{\,-s} Y_{jm},
\qquad s\le0 .
\end{align}
where the operators act algebraically according to
\begin{align}
\eth\,{}_sY_{jm}
&=
\sqrt{(j-s)(j+s+1)}\,
{}_{s+1}Y_{jm},
\\
\ethbar\,{}_sY_{jm}
&=
-\sqrt{(j+s)(j-s+1)}\,
{}_{s-1}Y_{jm},
\end{align}
and so satisfy
\begin{equation}
\ethbar\,\eth\,{}_sY_{jm}
=
-(j-s)(j+s+1)\,{}_sY_{jm}.
\end{equation}

Because the monopole background carries magnetic flux, angular derivatives acting on
charged fields must be gauge covariant. In the present case, the angular dependence is
expanded in monopole harmonics \cite{Gervalle:2022npx}, which coincide with spin--weighted spherical harmonics
once the background gauge connection is incorporated.
Operationally, this means that the angular part of the covariant Laplacian acting on
perturbations reduces to algebraic $j$-dependent factors after expansion in
${}_sY_{jm}$. This property underlies the complete separation of variables and the
reduction of the fluctuation equations to radial systems labelled by the total angular
momentum quantum number $j$.

After expansion in spin-weighted harmonics, perturbations decompose into independent
\emph{fixed-$j$ sectors}. Each sector contains a finite number of radial amplitudes, corresponding
to the different spin weights and internal indices of the fields.

In particular, for the hypercharge gauge field the physical transverse modes reside in the
$s=\pm1$ components, yielding two radial channels for each $j\ge1$. Components with $s=0$
belong to the time--radial sector and are eliminated by gauge fixing and constraint equations,
as discussed in the main text. This channel counting agrees with that of
Gervalle and Volkov and ensures that no spurious degrees of freedom are introduced.

\section{Gauge fixing and constraint elimination}
\label{app:gauge}

 Gauge fixing and constraint equations are used to eliminate
non-dynamical components of the gauge-field perturbations and to reduce the fluctuation
problem to a symmetric self-adjoint radial eigenvalue problem.
The construction follows the background-gauge strategy of Gervalle and Volkov \cite{Gervalle:2022npx,Gervalle:2022vxs}, specialised
here to the Abelian hypercharge sector and extended to include the Born--Infeld constitutive
relation.

The hypercharge gauge field enjoys the Abelian gauge symmetry
\begin{equation}
\delta B_\mu \;\rightarrow\; \delta B_\mu + \partial_\mu \chi ,
\end{equation}
where $\chi$ is an arbitrary scalar function.
To remove pure-gauge components and obtain a well-defined quadratic fluctuation operator,
we impose a Lorentz-type background gauge condition,
\begin{equation}
\nabla_\mu \delta B^\mu = 0 .
\label{eq:lorenzgauge}
\end{equation}
In contrast to non-Abelian background gauges, the derivative appearing in
\eqref{eq:lorenzgauge} is an ordinary spacetime divergence, since the hypercharge field
transforms trivially under its own gauge group.
This gauge choice fixes the residual $U(1)_Y$ gauge freedom and ensures that the fluctuation
operator is formally self-adjoint with respect to the natural inner product on the space of
perturbations.

In order to study fluctuations we linearise the Born--Infeld hypercharge equation of motion
\begin{equation}
\nabla_\mu G_Y^{\mu\nu} = J_Y^\nu
\end{equation}
about a fixed background yields
\begin{equation}
\nabla_\mu \delta G_Y^{\mu\nu} = \delta J_Y^\nu .
\label{eq:linEOM}
\end{equation}
In the hypercharge-only block considered in the main text, the \emph{coupled Higgs and
$SU(2)$ fluctuations are set to zero}, and the linearised Higgs constraint implies
$\delta J_Y^\nu = 0$.
The fluctuation equations therefore reduce to
\begin{equation}
\nabla_\mu \delta G_Y^{\mu\nu} = 0 .
\label{eq:sourcefree}
\end{equation}
The $\nu = t$ component of \eqref{eq:sourcefree} is Gauss' law.
It contains no second time derivatives of the perturbations and therefore represents a
\emph{constraint} rather than a dynamical equation.
Together with the gauge condition \eqref{eq:lorenzgauge}, Gauss' law determines the
time component $\delta B_t$ and the longitudinal part of the spatial perturbation
algebraically in terms of the transverse components.
As a result, $\delta B_t$ and the longitudinal mode do not represent independent degrees
of freedom and may be consistently eliminated.

In the complex null tetrad introduced in Appendix~\ref{app:angular}, the perturbation
$\delta B_\mu$ decomposes into components with spin weights
$s=0$ (along $l^\mu$ and $n^\mu$) and $s=\pm1$ (along $m^\mu$ and $\bar m^\mu$).
The $s=0$ components correspond to the \emph{time-radial} sector $(t,r)$ and enter the quadratic
action without kinetic terms.
They are therefore non-dynamical and are removed by solving the constraint equations
described above.

Importantly, the Born--Infeld modification does not alter this conclusion.
Although the constitutive relation modifies the algebraic relation between $\delta G_Y^{\mu\nu}$
and $\delta B_{\mu\nu}$, the linearised equations remain second order and retain their
divergence form.
Consequently, the constraint structure of the hypercharge sector is identical to that of
Maxwell theory, and no additional propagating degrees of freedom are introduced.

After gauge fixing and constraint elimination, the remaining physical perturbations reside
entirely in the transverse $s=\pm1$ components.
Expanding these components in spin--weighted spherical harmonics yields two independent
radial amplitudes for each fixed angular momentum $j\ge1$,
\begin{equation}
\Psi_j(r) =
\begin{pmatrix}
b_j^{+}(r) \\
b_j^{-}(r)
\end{pmatrix}.
\end{equation}
This channel counting coincides exactly with that obtained in the Maxwell theory and with
the analysis of Gervalle and Volkov, confirming that the Born--Infeld modification does not
change the number of physical modes.

Substituting the reduced ansatz into the linearised equations
\eqref{eq:sourcefree} and projecting onto a fixed-$j$ sector leads to a coupled radial
system of the Sturm--Liouville type,
\begin{equation}
-\frac{d}{dr}\!\left(P_j(r)\frac{d}{dr}\Psi_j\right)
+ Q_j(r)\,\Psi_j
=
\omega^2\,P_j(r)\,\Psi_j .
\label{eq:SLapp}
\end{equation}
Here $P_j(r)=-\bar{\mathcal L}_{X_Y}(r)>0$ is determined by the background Born--Infeld
response function, while $Q_j(r)$ is a symmetric matrix encoding centrifugal and
Born--Infeld-induced potential terms.
Because the fluctuation operator arises from a gauge-fixed quadratic action and the
constraint variables have been eliminated, the operator in \eqref{eq:SLapp} is manifestly
self-adjoint with respect to the inner product
\begin{equation}
\langle \Psi_1,\Psi_2\rangle
=
\int_0^\infty dr\, \Psi_1^\dagger(r)\,P_j(r)\,\Psi_2(r).
\end{equation}
This completes the justification of the reduction used in the main text and shows that the
Born--Infeld hypercharge sector admits a well-defined spectral problem with the same
constraint structure and channel counting as in the Maxwell case \cite{Gervalle:2022npx}.

\section{Sturm-Liouville structure, endpoint behaviour, and limitations}
\label{app:SL}

After analysing the radial Sturm-Liouville problem obtained from gauge fixing and
constraint elimination, we discuss the behaviour at the singular endpoint $r=0$ of the differential equation, and clarify
the role and limitations of the phenomenological regularisation employed in the main text.
Our purpose is to make explicit which aspects of the stability analysis are rigorous and
which rely on additional assumptions about the monopole core.

The reduced fixed-$j$ hypercharge fluctuation operator contains, in addition to the Maxwell-like
centrifugal term, an extra Born--Infeld contribution coming from the
$\bar{\mathcal L}_{X_YX_Y}$ term in the linearised constitutive relation
\eqref{eq:linconstitutive_app}. Accordingly the correct Sturm--Liouville system reads
\begin{equation}
-\frac{d}{dr}\!\left(P(r)\frac{d}{dr}\Psi_j\right)
+\Big(Q^{(\mathrm{Max})}_{Y,j}(r)+Q^{(\mathrm{BI})}_{Y,j}(r)\Big)\Psi_j
=\omega^2 P(r)\Psi_j ,
\end{equation}
with $P(r)=-\bar{\mathcal L}_{X_Y}(r)>0$ and $Q^{(\mathrm{BI})}_{Y,j}(r)$ a symmetric potential-type
matrix determined by the background functions $\bar B_{\mu\nu}(r)$, $\bar{\mathcal L}_{X_Y}(r)$ and
$\bar{\mathcal L}_{X_YX_Y}(r)$. The variational functional and any subsequent bounds must therefore
be formulated in terms of the full $Q_{Y,j}=Q^{(\mathrm{Max})}_{Y,j}+Q^{(\mathrm{BI})}_{Y,j}$.
Where an explicit background profile is not available, our analysis should be interpreted as a
truncated (Maxwell-like) estimate and a complete spectral determination requires numerical
evaluation of $Q^{(\mathrm{BI})}_{Y,j}(r)$.
To remove the variable weight, one may perform a Liouville-type transformation.
In the hypercharge-only reduction one has $P_j(r)=P(r)\,\mathbf 1_2$ with $P(r)>0$, so we set
\begin{equation}
\Psi_j(r)=P(r)^{-1/2}\,\Phi_j(r).
\label{eq:LiouvillePsiPhi}
\end{equation}
On writing the full potential as
\begin{equation}
Q_j(r)=Q^{(\mathrm{Max})}_{Y,j}(r)+Q^{(\mathrm{BI})}_{Y,j}(r),
\label{eq:Qsplit}
\end{equation}
(where $Q^{(\mathrm{BI})}_{Y,j}$ is the additional symmetric matrix induced by the
$\bar{\mathcal L}_{X_YX_Y}$ term in the linearised constitutive relation) the
matrix Sturm--Liouville system
\begin{equation}
-\frac{d}{dr}\!\left(P(r)\frac{d}{dr}\Psi_j\right)+Q_j(r)\Psi_j
=\omega^2 P(r)\Psi_j
\label{eq:SLmatrix_full}
\end{equation}
is transformed into a Schr\"odinger-type eigenvalue problem
\begin{equation}
-\Phi_j'' + V^{\rm eff}_j(r)\,\Phi_j=\omega^2\,\Phi_j,
\label{eq:Schro_full}
\end{equation}
with effective matrix potential
\begin{equation}
V^{\rm eff}_j(r)
=
P(r)^{-1/2}\,Q_j(r)\,P(r)^{-1/2}
+\frac{1}{2}\frac{P''(r)}{P(r)}\,\mathbf 1_2
-\frac{3}{4}\left(\frac{P'(r)}{P(r)}\right)^2\mathbf 1_2.
\label{eq:Veff_full}
\end{equation}
Here the derivative terms are proportional to the identity because $P(r)$ is a scalar weight
in the hypercharge block. The transformation is formal when $P(r)$ is smooth and strictly
positive; in particular, its applicability near $r=0$ depends on the endpoint behaviour of
the background response function $P(r)$ (and hence on the assumed short-distance completion
of the monopole core).

For a Dirac-type monopole background, the hypercharge magnetic field behaves as
$\bar B_Y\sim g/r^2$ near the origin, implying $\bar X_Y\sim g^2/(2r^4)$.
In the Maxwell theory this leads to $P(r)=1$, while in Born-Infeld theory one finds
\begin{equation}
P(r)\sim \frac{\beta r^2}{|g|},
\qquad r\to0 ,
\end{equation}
up to numerical factors depending on conventions.As a result, the coefficients in \eqref{eq:SLmatrix_full} become singular at $r=0$.
The endpoint $r=0$ is therefore of singular Sturm-Liouville type, and standard theorems
guaranteeing essential self-adjointness or positivity do not apply without further input.

To probe the effect of Born-Infeld smoothing on the fluctuation spectrum, the main text
introduces a phenomenological regularisation of the monopole core, for example by replacing
\begin{equation}
\bar B_Y(r)=\frac{g}{r^2}
\quad\longrightarrow\quad
\bar B_Y^{\rm reg}(r)=\frac{g}{r^2+a^2},
\end{equation}
with $a$ a short-distance cutoff.
This renders $P(r)$ and $Q_j(r)$ finite at the origin and converts the problem into a
regular Sturm--Liouville system on $[0,\infty)$ with well-defined boundary conditions.
For such regularised backgrounds, the operator in \eqref{eq:SLmatrix_full} is manifestly
self-adjoint and admits a variational characterisation of the spectrum.

For regularised backgrounds, the lowest eigenvalue satisfies
\begin{equation}
\omega_0^2
=
\inf_{\Psi\neq0}
\frac{
\displaystyle
\int_0^\infty dr\,
\big[
(P\,\Psi')^\dagger\Psi'
+ \Psi^\dagger Q_j \Psi
\big]
}{
\displaystyle
\int_0^\infty dr\, \Psi^\dagger P \Psi
}.
\end{equation}
Positivity of this functional for all admissible trial functions implies the absence of
negative modes and hence linear stability in the corresponding sector.
The results reported in the main text should be interpreted in this variational sense.

In the Born-Infeld case the reduced fixed-$j$ hypercharge potential splits as
\begin{equation}
Q_{Y,j}(r)=Q^{(\mathrm{Max})}_{Y,j}(r)+Q^{(\mathrm{BI})}_{Y,j}(r),
\end{equation}
where $Q^{(\mathrm{BI})}_{Y,j}$ arises from the $\bar{\mathcal L}_{X_YX_Y}$ term in the
linearised constitutive relation \eqref{eq:linconstitutive_app}.
For a static purely magnetic background one has
$\bar B_{\rho\sigma}\delta B^{\rho\sigma}=2\,\bar{\mathbf B}_Y\!\cdot\!\delta\mathbf B_Y$,
so after gauge/constraint elimination and projection to the transverse $s=\pm1$ sector the
Born--Infeld contribution takes the universal form
\begin{equation}
Q^{(\mathrm{BI})}_{Y,j}(r)
=
\bar{\mathcal L}_{X_YX_Y}(r)\,\bar{\mathbf B}_Y(r)^2\;\mathsf{C}_j(r),
\label{eq:QBI_struct}
\end{equation}
with $\mathsf{C}_j(r)$ a real symmetric $2\times2$ matrix fixed by the angular projection
and the chosen transverse basis.
In the helicity basis adapted to spherical symmetry $\mathsf{C}_j$ is proportional to the
identity, $\mathsf{C}_j=c_j(r)\mathbf 1_2$, but the overall sign of $Q^{(\mathrm{BI})}_{Y,j}$
is not determined a priori without an explicit background profile and normalisation
conventions for the reduced amplitudes.
Accordingly, stability criteria based on positivity of the reduced potential must either
include $Q^{(\mathrm{BI})}_{Y,j}$ explicitly (given a specified background) or be formulated
as bounds that allow for an indefinite symmetric contribution.

We emphasise that the above conclusions rely on two assumptions:
(i) the use of a phenomenologically regularised hypercharge background, and
(ii) the neglect of backreaction of the Born--Infeld modification on the Higgs and
non-Abelian gauge profiles.
While Born--Infeld dynamics is expected to soften the monopole core and localise its effects,
a fully rigorous stability analysis requires the numerical construction of the complete
coupled Born--Infeld electroweak monopole background followed by a spectral analysis of
fluctuations about that solution.
The present work should therefore be regarded as providing evidence for stability of the
hypercharge sector rather than a definitive proof for the full theory.

\bibliographystyle{apsrev4-2}

\bibliography{MonStab} 

\begin{thebibliography}{31}%
\makeatletter
\providecommand \@ifxundefined [1]{%
 \@ifx{#1\undefined}
}%
\providecommand \@ifnum [1]{%
 \ifnum #1\expandafter \@firstoftwo
 \else \expandafter \@secondoftwo
 \fi
}%
\providecommand \@ifx [1]{%
 \ifx #1\expandafter \@firstoftwo
 \else \expandafter \@secondoftwo
 \fi
}%
\providecommand \natexlab [1]{#1}%
\providecommand \enquote  [1]{``#1''}%
\providecommand \bibnamefont  [1]{#1}%
\providecommand \bibfnamefont [1]{#1}%
\providecommand \citenamefont [1]{#1}%
\providecommand \href@noop [0]{\@secondoftwo}%
\providecommand \href [0]{\begingroup \@sanitize@url \@href}%
\providecommand \@href[1]{\@@startlink{#1}\@@href}%
\providecommand \@@href[1]{\endgroup#1\@@endlink}%
\providecommand \@sanitize@url [0]{\catcode `\\12\catcode `\$12\catcode `\&12\catcode `\#12\catcode `\^12\catcode `\_12\catcode `\%12\relax}%
\providecommand \@@startlink[1]{}%
\providecommand \@@endlink[0]{}%
\providecommand \url  [0]{\begingroup\@sanitize@url \@url }%
\providecommand \@url [1]{\endgroup\@href {#1}{\urlprefix }}%
\providecommand \urlprefix  [0]{URL }%
\providecommand \Eprint [0]{\href }%
\providecommand \doibase [0]{https://doi.org/}%
\providecommand \selectlanguage [0]{\@gobble}%
\providecommand \bibinfo  [0]{\@secondoftwo}%
\providecommand \bibfield  [0]{\@secondoftwo}%
\providecommand \translation [1]{[#1]}%
\providecommand \BibitemOpen [0]{}%
\providecommand \bibitemStop [0]{}%
\providecommand \bibitemNoStop [0]{.\EOS\space}%
\providecommand \EOS [0]{\spacefactor3000\relax}%
\providecommand \BibitemShut  [1]{\csname bibitem#1\endcsname}%
\let\auto@bib@innerbib\@empty
\bibitem [{\citenamefont {Dirac}(1931)}]{Dirac1931}%
  \BibitemOpen
  \bibfield  {author} {\bibinfo {author} {\bibfnamefont {P.~A.~M.}\ \bibnamefont {Dirac}},\ }\href {https://doi.org/10.1098/rspa.1931.0130} {\bibfield  {journal} {\bibinfo  {journal} {Proc. Roy. Soc. Lond. A}\ }\textbf {\bibinfo {volume} {133}},\ \bibinfo {pages} {60} (\bibinfo {year} {1931})}\BibitemShut {NoStop}%
\bibitem [{\citenamefont {Dirac}(1948)}]{Dirac:1948um}%
  \BibitemOpen
  \bibfield  {author} {\bibinfo {author} {\bibfnamefont {P.~A.~M.}\ \bibnamefont {Dirac}},\ }\href {https://doi.org/10.1103/PhysRev.74.817} {\bibfield  {journal} {\bibinfo  {journal} {Phys. Rev.}\ }\textbf {\bibinfo {volume} {74}},\ \bibinfo {pages} {817} (\bibinfo {year} {1948})}\BibitemShut {NoStop}%
\bibitem [{\citenamefont {'t~Hooft}(1974)}]{tHooft1974}%
  \BibitemOpen
  \bibfield  {author} {\bibinfo {author} {\bibfnamefont {G.}~\bibnamefont {'t~Hooft}},\ }\href {https://doi.org/10.1016/0550-3213(74)90486-6} {\bibfield  {journal} {\bibinfo  {journal} {Nucl. Phys. B}\ }\textbf {\bibinfo {volume} {79}},\ \bibinfo {pages} {276} (\bibinfo {year} {1974})}\BibitemShut {NoStop}%
\bibitem [{\citenamefont {Polyakov}(1974)}]{Polyakov1974}%
  \BibitemOpen
  \bibfield  {author} {\bibinfo {author} {\bibfnamefont {A.~M.}\ \bibnamefont {Polyakov}},\ }\href@noop {} {\bibfield  {journal} {\bibinfo  {journal} {JETP Lett.}\ }\textbf {\bibinfo {volume} {20}},\ \bibinfo {pages} {194} (\bibinfo {year} {1974})}\BibitemShut {NoStop}%
\bibitem [{\citenamefont {Weinberg}(1967)}]{Weinberg1967}%
  \BibitemOpen
  \bibfield  {author} {\bibinfo {author} {\bibfnamefont {S.}~\bibnamefont {Weinberg}},\ }\href {https://doi.org/10.1103/PhysRevLett.19.1264} {\bibfield  {journal} {\bibinfo  {journal} {Phys. Rev. Lett.}\ }\textbf {\bibinfo {volume} {19}},\ \bibinfo {pages} {1264} (\bibinfo {year} {1967})}\BibitemShut {NoStop}%
\bibitem [{\citenamefont {Salam}(1968)}]{Salam1968}%
  \BibitemOpen
  \bibfield  {author} {\bibinfo {author} {\bibfnamefont {A.}~\bibnamefont {Salam}},\ }in\ \href@noop {} {\emph {\bibinfo {booktitle} {Elementary Particle Theory: Relativistic Groups and Analyticity}}},\ \bibinfo {editor} {edited by\ \bibinfo {editor} {\bibfnamefont {N.}~\bibnamefont {Svartholm}}}\ (\bibinfo  {publisher} {Almqvist and Wiksell},\ \bibinfo {address} {Stockholm},\ \bibinfo {year} {1968})\ pp.\ \bibinfo {pages} {367--377}\BibitemShut {NoStop}%
\bibitem [{\citenamefont {Cho}\ and\ \citenamefont {Maison}(1997)}]{ChoMaison1997}%
  \BibitemOpen
  \bibfield  {author} {\bibinfo {author} {\bibfnamefont {Y.~M.}\ \bibnamefont {Cho}}\ and\ \bibinfo {author} {\bibfnamefont {D.}~\bibnamefont {Maison}},\ }\href {https://doi.org/10.1016/S0370-2693(96)01492-X} {\bibfield  {journal} {\bibinfo  {journal} {Phys. Lett. B}\ }\textbf {\bibinfo {volume} {391}},\ \bibinfo {pages} {360} (\bibinfo {year} {1997})},\ \Eprint {https://arxiv.org/abs/hep-th/9601028} {arXiv:hep-th/9601028} \BibitemShut {NoStop}%
\bibitem [{\citenamefont {Fradkin}\ and\ \citenamefont {Tseytlin}(1985)}]{Fradkin:1985qd}%
  \BibitemOpen
  \bibfield  {author} {\bibinfo {author} {\bibfnamefont {E.~S.}\ \bibnamefont {Fradkin}}\ and\ \bibinfo {author} {\bibfnamefont {A.~A.}\ \bibnamefont {Tseytlin}},\ }\href {https://doi.org/10.1016/0370-2693(85)91599-0} {\bibfield  {journal} {\bibinfo  {journal} {Phys. Lett. B}\ }\textbf {\bibinfo {volume} {163}},\ \bibinfo {pages} {123} (\bibinfo {year} {1985})}\BibitemShut {NoStop}%
\bibitem [{\citenamefont {Leigh}(1989)}]{Leigh:1989jq}%
  \BibitemOpen
  \bibfield  {author} {\bibinfo {author} {\bibfnamefont {R.~G.}\ \bibnamefont {Leigh}},\ }\href {https://doi.org/10.1016/0315-591X(89)90015-8} {\bibfield  {journal} {\bibinfo  {journal} {Mod. Phys. Lett. A}\ }\textbf {\bibinfo {volume} {4}},\ \bibinfo {pages} {2767} (\bibinfo {year} {1989})}\BibitemShut {NoStop}%
\bibitem [{\citenamefont {Gervalle}\ and\ \citenamefont {Volkov}(2022)}]{Gervalle:2022npx}%
  \BibitemOpen
  \bibfield  {author} {\bibinfo {author} {\bibfnamefont {R.}~\bibnamefont {Gervalle}}\ and\ \bibinfo {author} {\bibfnamefont {M.~S.}\ \bibnamefont {Volkov}},\ }\href {https://doi.org/10.1016/j.nuclphysb.2022.115937} {\bibfield  {journal} {\bibinfo  {journal} {Nucl. Phys. B}\ }\textbf {\bibinfo {volume} {984}},\ \bibinfo {pages} {115937} (\bibinfo {year} {2022})},\ \Eprint {https://arxiv.org/abs/2203.16590} {arXiv:2203.16590 [hep-th]} \BibitemShut {NoStop}%
\bibitem [{\citenamefont {Gervalle}\ and\ \citenamefont {Volkov}(2023)}]{Gervalle:2022vxs}%
  \BibitemOpen
  \bibfield  {author} {\bibinfo {author} {\bibfnamefont {R.}~\bibnamefont {Gervalle}}\ and\ \bibinfo {author} {\bibfnamefont {M.~S.}\ \bibnamefont {Volkov}},\ }\href {https://doi.org/10.1016/j.nuclphysb.2023.116112} {\bibfield  {journal} {\bibinfo  {journal} {Nucl. Phys. B}\ }\textbf {\bibinfo {volume} {987}},\ \bibinfo {pages} {116112} (\bibinfo {year} {2023})},\ \Eprint {https://arxiv.org/abs/2211.04875} {arXiv:2211.04875 [hep-th]} \BibitemShut {NoStop}%
\bibitem [{\citenamefont {Arunasalam}\ and\ \citenamefont {Kobakhidze}(2017)}]{ArunasalamKobakhidze2017}%
  \BibitemOpen
  \bibfield  {author} {\bibinfo {author} {\bibfnamefont {S.}~\bibnamefont {Arunasalam}}\ and\ \bibinfo {author} {\bibfnamefont {A.}~\bibnamefont {Kobakhidze}},\ }\href {https://doi.org/10.1140/epjc/s10052-017-4999-y} {\bibfield  {journal} {\bibinfo  {journal} {Eur. Phys. J. C}\ }\textbf {\bibinfo {volume} {77}},\ \bibinfo {pages} {444} (\bibinfo {year} {2017})},\ \Eprint {https://arxiv.org/abs/1702.04068} {arXiv:1702.04068 [hep-ph]} \BibitemShut {NoStop}%
\bibitem [{\citenamefont {Mavromatos}\ and\ \citenamefont {Sarkar}(2018)}]{MavromatosSarkar2019}%
  \BibitemOpen
  \bibfield  {author} {\bibinfo {author} {\bibfnamefont {N.~E.}\ \bibnamefont {Mavromatos}}\ and\ \bibinfo {author} {\bibfnamefont {S.}~\bibnamefont {Sarkar}},\ }\href {https://doi.org/10.3390/universe5010008} {\bibfield  {journal} {\bibinfo  {journal} {Universe}\ }\textbf {\bibinfo {volume} {5}},\ \bibinfo {pages} {8} (\bibinfo {year} {2018})},\ \Eprint {https://arxiv.org/abs/1812.00495} {arXiv:1812.00495 [hep-ph]} \BibitemShut {NoStop}%
\bibitem [{\citenamefont {Kim}(2000)}]{Kim2000}%
  \BibitemOpen
  \bibfield  {author} {\bibinfo {author} {\bibfnamefont {H.}~\bibnamefont {Kim}},\ }\href {https://doi.org/10.1103/PhysRevD.61.085014} {\bibfield  {journal} {\bibinfo  {journal} {Phys. Rev. D}\ }\textbf {\bibinfo {volume} {61}},\ \bibinfo {pages} {085014} (\bibinfo {year} {2000})},\ \Eprint {https://arxiv.org/abs/hep-th/9910261} {arXiv:hep-th/9910261} \BibitemShut {NoStop}%
\bibitem [{\citenamefont {Born}\ and\ \citenamefont {Infeld}(1934)}]{Born:1934gh}%
  \BibitemOpen
  \bibfield  {author} {\bibinfo {author} {\bibfnamefont {M.}~\bibnamefont {Born}}\ and\ \bibinfo {author} {\bibfnamefont {L.}~\bibnamefont {Infeld}},\ }\href {https://doi.org/10.1098/rspa.1934.0059} {\bibfield  {journal} {\bibinfo  {journal} {Proc. Roy. Soc. Lond. A}\ }\textbf {\bibinfo {volume} {144}},\ \bibinfo {pages} {425} (\bibinfo {year} {1934})}\BibitemShut {NoStop}%
\bibitem [{\citenamefont {Ellis}\ \emph {et~al.}(2017)\citenamefont {Ellis}, \citenamefont {Mavromatos},\ and\ \citenamefont {You}}]{Ellis2017LbL}%
  \BibitemOpen
  \bibfield  {author} {\bibinfo {author} {\bibfnamefont {J.}~\bibnamefont {Ellis}}, \bibinfo {author} {\bibfnamefont {N.~E.}\ \bibnamefont {Mavromatos}},\ and\ \bibinfo {author} {\bibfnamefont {T.}~\bibnamefont {You}},\ }\href {https://doi.org/10.1103/PhysRevLett.118.261802} {\bibfield  {journal} {\bibinfo  {journal} {Phys. Rev. Lett.}\ }\textbf {\bibinfo {volume} {118}},\ \bibinfo {pages} {261802} (\bibinfo {year} {2017})},\ \Eprint {https://arxiv.org/abs/1703.08450} {arXiv:1703.08450 [hep-ph]} \BibitemShut {NoStop}%
\bibitem [{\citenamefont {Grandi}\ \emph {et~al.}(1999)\citenamefont {Grandi}, \citenamefont {Pakman}, \citenamefont {Schaposnik},\ and\ \citenamefont {Silva}}]{Grandi1999}%
  \BibitemOpen
  \bibfield  {author} {\bibinfo {author} {\bibfnamefont {N.~E.}\ \bibnamefont {Grandi}}, \bibinfo {author} {\bibfnamefont {A.}~\bibnamefont {Pakman}}, \bibinfo {author} {\bibfnamefont {F.~A.}\ \bibnamefont {Schaposnik}},\ and\ \bibinfo {author} {\bibfnamefont {G.~A.}\ \bibnamefont {Silva}},\ }\href {https://doi.org/10.1103/PhysRevD.60.125002} {\bibfield  {journal} {\bibinfo  {journal} {Phys. Rev. D}\ }\textbf {\bibinfo {volume} {60}},\ \bibinfo {pages} {125002} (\bibinfo {year} {1999})},\ \Eprint {https://arxiv.org/abs/hep-th/9906244} {arXiv:hep-th/9906244} \BibitemShut {NoStop}%
\bibitem [{\citenamefont {Ellis}\ \emph {et~al.}(2022)\citenamefont {Ellis}, \citenamefont {Mavromatos}, \citenamefont {Roloff},\ and\ \citenamefont {You}}]{Ellis:2022uxv}%
  \BibitemOpen
  \bibfield  {author} {\bibinfo {author} {\bibfnamefont {J.}~\bibnamefont {Ellis}}, \bibinfo {author} {\bibfnamefont {N.~E.}\ \bibnamefont {Mavromatos}}, \bibinfo {author} {\bibfnamefont {P.}~\bibnamefont {Roloff}},\ and\ \bibinfo {author} {\bibfnamefont {T.}~\bibnamefont {You}},\ }\href {https://doi.org/10.1140/epjc/s10052-022-10565-w} {\bibfield  {journal} {\bibinfo  {journal} {Eur. Phys. J. C}\ }\textbf {\bibinfo {volume} {82}},\ \bibinfo {pages} {634} (\bibinfo {year} {2022})},\ \Eprint {https://arxiv.org/abs/2203.17111} {arXiv:2203.17111 [hep-ph]} \BibitemShut {NoStop}%
\bibitem [{\citenamefont {Aad}\ \emph {et~al.}(2023)\citenamefont {Aad} \emph {et~al.}}]{ATLAS:2023esy}%
  \BibitemOpen
  \bibfield  {author} {\bibinfo {author} {\bibfnamefont {G.}~\bibnamefont {Aad}} \emph {et~al.} (\bibinfo {collaboration} {ATLAS}),\ }\href {https://doi.org/10.1007/JHEP11(2023)112} {\bibfield  {journal} {\bibinfo  {journal} {JHEP}\ }\textbf {\bibinfo {volume} {11}},\ \bibinfo {pages} {112}},\ \Eprint {https://arxiv.org/abs/2308.04835} {arXiv:2308.04835 [hep-ex]} \BibitemShut {NoStop}%
\bibitem [{\citenamefont {Aad}\ \emph {et~al.}(2025)\citenamefont {Aad} \emph {et~al.}}]{ATLAS:2024nzp}%
  \BibitemOpen
  \bibfield  {author} {\bibinfo {author} {\bibfnamefont {G.}~\bibnamefont {Aad}} \emph {et~al.} (\bibinfo {collaboration} {ATLAS}),\ }\href {https://doi.org/10.1103/PhysRevLett.134.061803} {\bibfield  {journal} {\bibinfo  {journal} {Phys. Rev. Lett.}\ }\textbf {\bibinfo {volume} {134}},\ \bibinfo {pages} {061803} (\bibinfo {year} {2025})},\ \Eprint {https://arxiv.org/abs/2408.11035} {arXiv:2408.11035 [hep-ex]} \BibitemShut {NoStop}%
\bibitem [{\citenamefont {Acharya}\ \emph {et~al.}(2019)\citenamefont {Acharya} \emph {et~al.}}]{MoEDAL:2019ort}%
  \BibitemOpen
  \bibfield  {author} {\bibinfo {author} {\bibfnamefont {B.}~\bibnamefont {Acharya}} \emph {et~al.} (\bibinfo {collaboration} {MoEDAL}),\ }\href {https://doi.org/10.1103/PhysRevLett.123.021802} {\bibfield  {journal} {\bibinfo  {journal} {Phys. Rev. Lett.}\ }\textbf {\bibinfo {volume} {123}},\ \bibinfo {pages} {021802} (\bibinfo {year} {2019})},\ \Eprint {https://arxiv.org/abs/1903.08491} {arXiv:1903.08491 [hep-ex]} \BibitemShut {NoStop}%
\bibitem [{\citenamefont {Acharya}\ \emph {et~al.}(2022)\citenamefont {Acharya} \emph {et~al.}}]{MoEDAL:2021vix}%
  \BibitemOpen
  \bibfield  {author} {\bibinfo {author} {\bibfnamefont {B.}~\bibnamefont {Acharya}} \emph {et~al.} (\bibinfo {collaboration} {MoEDAL}),\ }\href {https://doi.org/10.1038/s41586-021-04298-1} {\bibfield  {journal} {\bibinfo  {journal} {Nature}\ }\textbf {\bibinfo {volume} {602}},\ \bibinfo {pages} {63} (\bibinfo {year} {2022})},\ \Eprint {https://arxiv.org/abs/2106.11933} {arXiv:2106.11933 [hep-ex]} \BibitemShut {NoStop}%
\bibitem [{\citenamefont {Acharya}\ \emph {et~al.}(2025)\citenamefont {Acharya} \emph {et~al.}}]{PhysRevLett.134.071802}%
  \BibitemOpen
  \bibfield  {author} {\bibinfo {author} {\bibfnamefont {B.}~\bibnamefont {Acharya}} \emph {et~al.} (\bibinfo {collaboration} {MoEDAL}),\ }\href {https://doi.org/10.1103/PhysRevLett.134.071802} {\bibfield  {journal} {\bibinfo  {journal} {Phys. Rev. Lett.}\ }\textbf {\bibinfo {volume} {134}},\ \bibinfo {pages} {071802} (\bibinfo {year} {2025})},\ \Eprint {https://arxiv.org/abs/2311.06509} {arXiv:2311.06509 [hep-ex]} \BibitemShut {NoStop}%
\bibitem [{\citenamefont {Laue}(1911)}]{Laue:1911lrk}%
  \BibitemOpen
  \bibfield  {author} {\bibinfo {author} {\bibfnamefont {M.}~\bibnamefont {Laue}},\ }\href {https://doi.org/10.1002/andp.19113400808} {\bibfield  {journal} {\bibinfo  {journal} {Annalen Phys.}\ }\textbf {\bibinfo {volume} {340}},\ \bibinfo {pages} {524} (\bibinfo {year} {1911})}\BibitemShut {NoStop}%
\bibitem [{\citenamefont {Farakos}\ \emph {et~al.}(2025)\citenamefont {Farakos}, \citenamefont {Koutsoumbas}, \citenamefont {Mavromatos},\ and\ \citenamefont {Zarafonitis}}]{Farakos:2025byy}%
  \BibitemOpen
  \bibfield  {author} {\bibinfo {author} {\bibfnamefont {K.}~\bibnamefont {Farakos}}, \bibinfo {author} {\bibfnamefont {G.}~\bibnamefont {Koutsoumbas}}, \bibinfo {author} {\bibfnamefont {N.~E.}\ \bibnamefont {Mavromatos}},\ and\ \bibinfo {author} {\bibfnamefont {A.}~\bibnamefont {Zarafonitis}},\ }\bibfield  {journal} {\bibinfo  {journal} {Eur. Phys. J. ST in press,}\ }\href {https://doi.org/10.1140/epjs/s11734-025-02083-z} {10.1140/epjs/s11734-025-02083-z} (\bibinfo {year} {2025}),\ \Eprint {https://arxiv.org/abs/2506.04872} {arXiv:2506.04872 [hep-th]} \BibitemShut {NoStop}%
\bibitem [{\citenamefont {Polyakov}(2003)}]{Polyakov:2002yz}%
  \BibitemOpen
  \bibfield  {author} {\bibinfo {author} {\bibfnamefont {M.~V.}\ \bibnamefont {Polyakov}},\ }\href {https://doi.org/10.1016/S0370-2693(03)00036-4} {\bibfield  {journal} {\bibinfo  {journal} {Phys. Lett. B}\ }\textbf {\bibinfo {volume} {555}},\ \bibinfo {pages} {57} (\bibinfo {year} {2003})},\ \Eprint {https://arxiv.org/abs/hep-ph/0210165} {arXiv:hep-ph/0210165} \BibitemShut {NoStop}%
\bibitem [{\citenamefont {Perevalova}\ \emph {et~al.}(2016)\citenamefont {Perevalova}, \citenamefont {Polyakov},\ and\ \citenamefont {Schweitzer}}]{Perevalova:2016dln}%
  \BibitemOpen
  \bibfield  {author} {\bibinfo {author} {\bibfnamefont {I.~A.}\ \bibnamefont {Perevalova}}, \bibinfo {author} {\bibfnamefont {M.~V.}\ \bibnamefont {Polyakov}},\ and\ \bibinfo {author} {\bibfnamefont {P.}~\bibnamefont {Schweitzer}},\ }\href {https://doi.org/10.1103/PhysRevD.94.054024} {\bibfield  {journal} {\bibinfo  {journal} {Phys. Rev. D}\ }\textbf {\bibinfo {volume} {94}},\ \bibinfo {pages} {054024} (\bibinfo {year} {2016})},\ \Eprint {https://arxiv.org/abs/1607.07008} {arXiv:1607.07008 [hep-ph]} \BibitemShut {NoStop}%
\bibitem [{\citenamefont {Panteleeva}(2023)}]{Panteleeva:2023aiz}%
  \BibitemOpen
  \bibfield  {author} {\bibinfo {author} {\bibfnamefont {J.~Y.}\ \bibnamefont {Panteleeva}},\ }\href {https://doi.org/10.1103/PhysRevD.107.055015} {\bibfield  {journal} {\bibinfo  {journal} {Phys. Rev. D}\ }\textbf {\bibinfo {volume} {107}},\ \bibinfo {pages} {055015} (\bibinfo {year} {2023})},\ \Eprint {https://arxiv.org/abs/2302.11980} {arXiv:2302.11980 [hep-ph]} \BibitemShut {NoStop}%
\bibitem [{\citenamefont {Peskin}\ and\ \citenamefont {Schroeder}(1995)}]{Peskin:1995ev}%
  \BibitemOpen
  \bibfield  {author} {\bibinfo {author} {\bibfnamefont {M.~E.}\ \bibnamefont {Peskin}}\ and\ \bibinfo {author} {\bibfnamefont {D.~V.}\ \bibnamefont {Schroeder}},\ }\href {https://doi.org/10.1201/9780429503559} {\emph {\bibinfo {title} {{An Introduction to quantum field theory}}}}\ (\bibinfo  {publisher} {Addison-Wesley},\ \bibinfo {address} {Reading, USA},\ \bibinfo {year} {1995})\BibitemShut {NoStop}%
\bibitem [{\citenamefont {d'Inverno}\ and\ \citenamefont {Vickers}(2022)}]{DInverno2022-cb}%
  \BibitemOpen
  \bibfield  {author} {\bibinfo {author} {\bibfnamefont {R.}~\bibnamefont {d'Inverno}}\ and\ \bibinfo {author} {\bibfnamefont {J.}~\bibnamefont {Vickers}},\ }\href@noop {} {\emph {\bibinfo {title} {Introducing Einstein's relativity}}},\ \bibinfo {edition} {2nd}\ ed.\ (\bibinfo  {publisher} {Oxford University Press},\ \bibinfo {address} {London, England},\ \bibinfo {year} {2022})\BibitemShut {NoStop}%
\bibitem [{\citenamefont {Shnir}(2005)}]{Shnir:2005vvi}%
  \BibitemOpen
  \bibfield  {author} {\bibinfo {author} {\bibfnamefont {Y.~M.}\ \bibnamefont {Shnir}},\ }\href {https://doi.org/10.1007/3-540-29082-6} {\emph {\bibinfo {title} {{Magnetic Monopoles}}}},\ Text and Monographs in Physics\ (\bibinfo  {publisher} {Springer},\ \bibinfo {address} {Berlin/Heidelberg},\ \bibinfo {year} {2005})\BibitemShut {NoStop}%
\end{thebibliography}%
\end{document}